\documentclass[pre, twocolumn, preprintnumbers, superscriptaddress]{revtex4-1}
\usepackage[]{graphicx}
\usepackage{epstopdf}
\usepackage{bm}
\usepackage{hyperref}
\usepackage{framed}
\usepackage{wasysym}
\usepackage{epstopdf}

\begin{document}
\title{Kinetic Spinodal Instabilities in the Mott Transition in V$_2$O$_3$: Evidence from Hysteresis Scaling and Dissipative Phase Ordering}
\author{Tapas Bar}
\affiliation{Indian Institute of Science Education \& Research Kolkata, Mohanpur, Nadia 741246, West Bengal, India}
\author{Sujeet Kumar Choudhary}\email{Current Affiliation: Department of Physics, Indian Institute of Science,  Bengaluru 560012, India}
\affiliation{Indian Institute of Science Education \& Research Kolkata, Mohanpur, Nadia 741246, West Bengal, India}
\author{Md. Arsalan Ashraf}\email{Current Affiliation: Raman Research Institute,  Bengaluru  560080, India}
\affiliation{Indian Institute of Science Education \& Research Kolkata, Mohanpur, Nadia 741246, West Bengal, India}
\author{K. S. Sujith} \email{Current Affiliation: School of Physical and Mathematical Sciences, Nanyang Technological University, 637371 Singapore}
\affiliation{Indian Institute of Science Education \& Research Kolkata, Mohanpur, Nadia 741246, West Bengal, India}
\author{Sanjay Puri} \affiliation{School of Physical Sciences, Jawaharlal Nehru University 110067, New Delhi, India}
\author{Satyabrata Raj}
\affiliation{Indian Institute of Science Education \& Research Kolkata, Mohanpur, Nadia 741246, West Bengal, India}
\author{Bhavtosh Bansal}\email{bhavtosh@iiserkol.ac.in}
\affiliation{Indian Institute of Science Education \& Research Kolkata, Mohanpur, Nadia 741246, West Bengal, India}
\date{May 29, 2018}
\begin{abstract}
We present the first systematic observation of scaling of thermal hysteresis with the temperature scanning rate around an abrupt thermodynamic transition in correlated electron systems. We show that the depth of supercooling and superheating in vanadium sesquioxide (V$_2$O$_3$) shifts with the temperature quench rates. The dynamic scaling exponent is close to the mean field prediction of 2/3. These observations, combined with the purely dissipative continuous ordering seen in ``quench-and-hold'' experiments, indicate departures from classical nucleation theory toward a barrier-free phase ordering associated with critical dynamics. Observation of critical-like features and scaling in a thermally induced abrupt phase transition suggests that the presence of a spinodal-like instability is not just an artifact of the mean field theories but can also exist in the transformation kinetics of real systems, surviving fluctuations.
\end{abstract}
\preprint{Physical Review Letters {\bf 121}, 045701 (2018)}
\maketitle
Metastable states do not exist in equilibrium statistical mechanics as any legitimate free energy must be convex in the thermodynamic limit \cite{binder_rpp}.  But many real systems do spontaneously fall out of equilibrium in a window of thermal hysteresis around the abrupt phase transition (APT) \cite{Debenedetti}.  The accompanying nonergodic behavior---arrested kinetics \cite{grygiel, chattopadhyay}, spatial inhomogeneity \cite{mcleod, Alsaqqa} and phase coexistence \cite{nandi, miao, liu}, and  rate dependence \cite{levy, Liu_simulation_rateDep, perez-Reche_rate}---is well documented.
\begin{figure}[!b]
\includegraphics[clip,width=9cm]{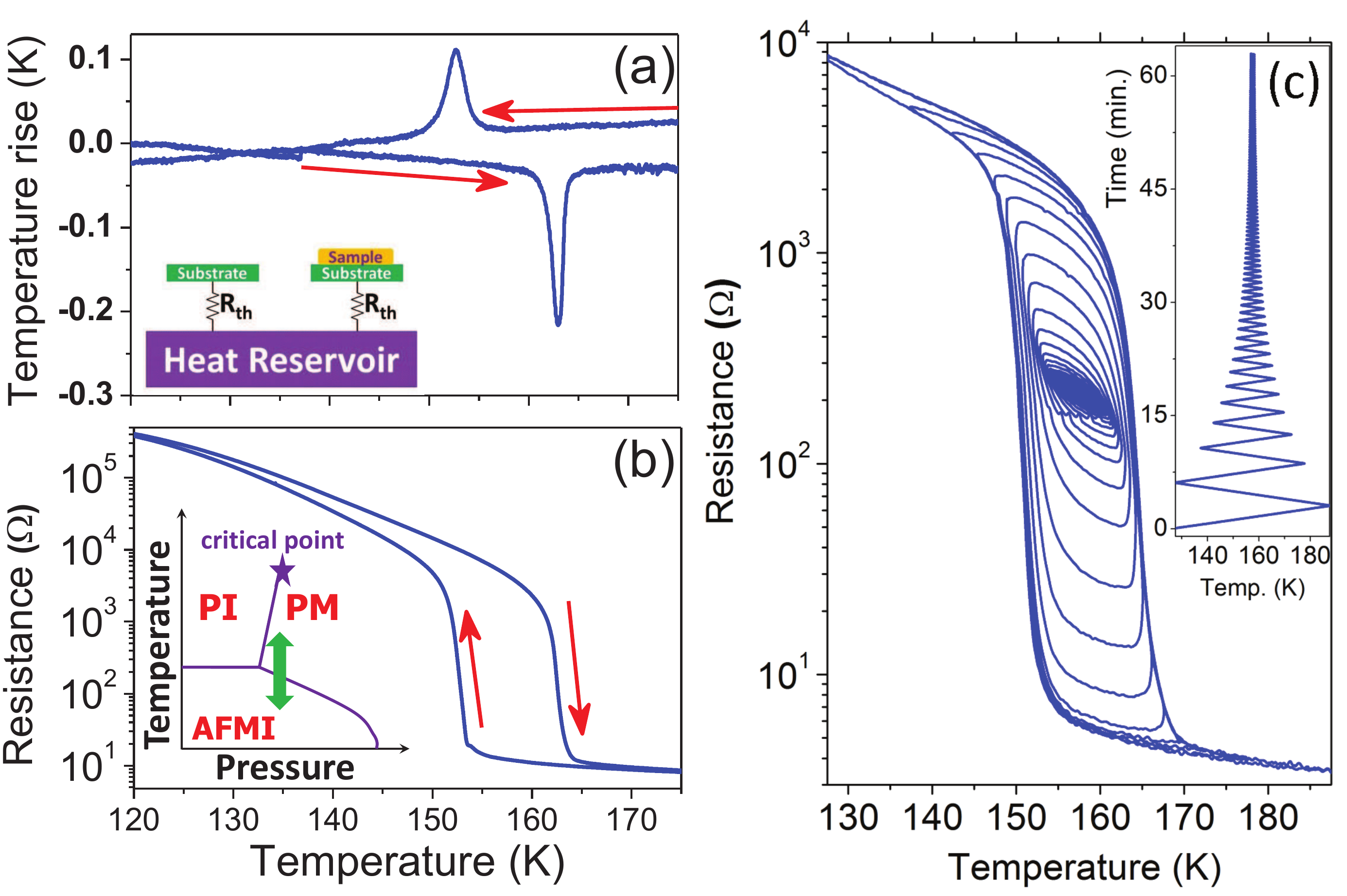}
\caption{(a) Exothermic ($\sim 153$ K) and endothermic ($\sim 162$ K)  latent heat peaks in the DTA experiment.  (Inset) Schematic of the set up \cite{SupplementaryMaterial}. (b) The resistance changes over many orders of magnitude around the same temperature. (a), (b) Quasiequilibrium properties measured at slow temperature ramp rates ($\leq 0.5$ K/min).  (Inset) Phase diagram of V$_2$O$_3$ in the pressure-temperature plane with antiferromagnetic insulator (AFMI),  paramagnetic insulator (PI), and paramagnetic metal (PM) phases separated by first-order lines \cite{McWhan_V2O3}. The arrow corresponds to the temperature-dependent transition (at ambient pressure) studied in this Letter. This transition also corresponds to a structural change from a rhombohedral (PM) to monoclinic phase (AFMI).  (c) Metastability of the hysteretic region is seen in the multivaluedness of the sample resistance. The fraction of the phase-transformed material within this region can be controlled by the sample's thermal history. (Inset) The corresponding time dependence of the temperature sweep. }  \label{Fig:Fig1}
\end{figure}

Within the mean field (MF) picture, this metastable phase is predicted to abruptly terminate at the spinodals, the two values of field or temperature where the barrier against nucleation vanishes \cite{Debenedetti, bray, Furukawa_Review, abaimov, Ivanov, nandi, Chu-Fisher, Gunton-Yalabik}. The analogy between the MF spinodals and the critical point in the power law divergence of susceptibility \cite{Chu-Fisher, Ivanov, Debenedetti, Saito, ikeda} and their being fixed points under renormalization group transformation \cite{zhong_pre2017, Gunton-Yalabik} has long been discussed \cite{Chu-Fisher, Debenedetti, footnote_speudospinodal}. Except for the case of strictly athermal systems \cite{Zapperi, abaimov, nandi}, these ideas were never taken seriously because one would expect this singularity to be physically inaccessible; fluctuations accompanying any finite-temperature phase transformation would necessarily yield pathways involving nucleation before the spinodal is experimentally reached \cite{ikeda, Ivanov, binder_rpp, Binder_puri-wadhawan}.

Nevertheless long-ranged forces arising, for example, due to the accompanying structural transition \cite{Rasmussen} may to an extent \cite{binder_ginzburgCriterion} suppress fluctuations. This will naturally lead to deep supersaturation and thermal hysteresis in the phase transformation, and thus take the system beyond the regime of the classical nucleation theory \cite{trudu,Maibaum, klein-monette, santra-bagchi, Gulbahce-Gould-Klein}. As the nucleation barriers get smaller, simulations show spatially diffused and continuous ordering mechanisms \cite{trudu} where the dynamic limit of metastability can extend to the critical nucleus shrinking to less than one molecule \cite{Maibaum}. Hence operationally, the phase ordering may actually mimic the MF spinodal behavior, with fluctuations only making quantitative corrections. In fact there is increasing theoretical evidence that the essence of this MF picture, i.e. the existence of singular fixed points, is retained in the dynamical behavior even for model systems with short-ranged interactions at finite temperature \cite{zhong_pre2017, Pelissetto}.

Focusing on the APT in V$_2$O$_3$  \cite{mcleod, McWhan_V2O3, imada_rmp, Rodolakis}, in this Letter we report the first experimental observation of such dressed MF behavior in phase ordering via the study of dynamic hysteresis.

\noindent
{\em Experiments.---} Figures 1(a) and 1(b) show quasiequilibrium differential thermal analysis (DTA) and transport measurements done at temperature ramp rates of $< 0.5$ K/min  using polycrystalline V$_2$O$_3$ (see Supplemental Material \cite{SupplementaryMaterial}); the transition is strongly hysteretic and the abrupt change in resistance is accompanied by latent heat $L$ $\approx 2$ kJ/mole ($\sim 15 RT_c$) \cite{keer}.  This APT is known to arise due to three simultaneous---electronic, structural, and magnetic---transformations [Fig. 1 (b) inset] \cite{McWhan_V2O3, imada_rmp}.  That the window of thermal hysteresis around $\sim 158$ K does correspond to the metastable region where the physical observables are no longer state variables is seen in the multivaluedness of the sample resistance in Fig. 1 (c). These minor hysteresis loops were drawn using the temperature protocol shown in Fig. 1 (c) (inset).
\begin{figure}[!t]
\includegraphics[clip,width=9cm]{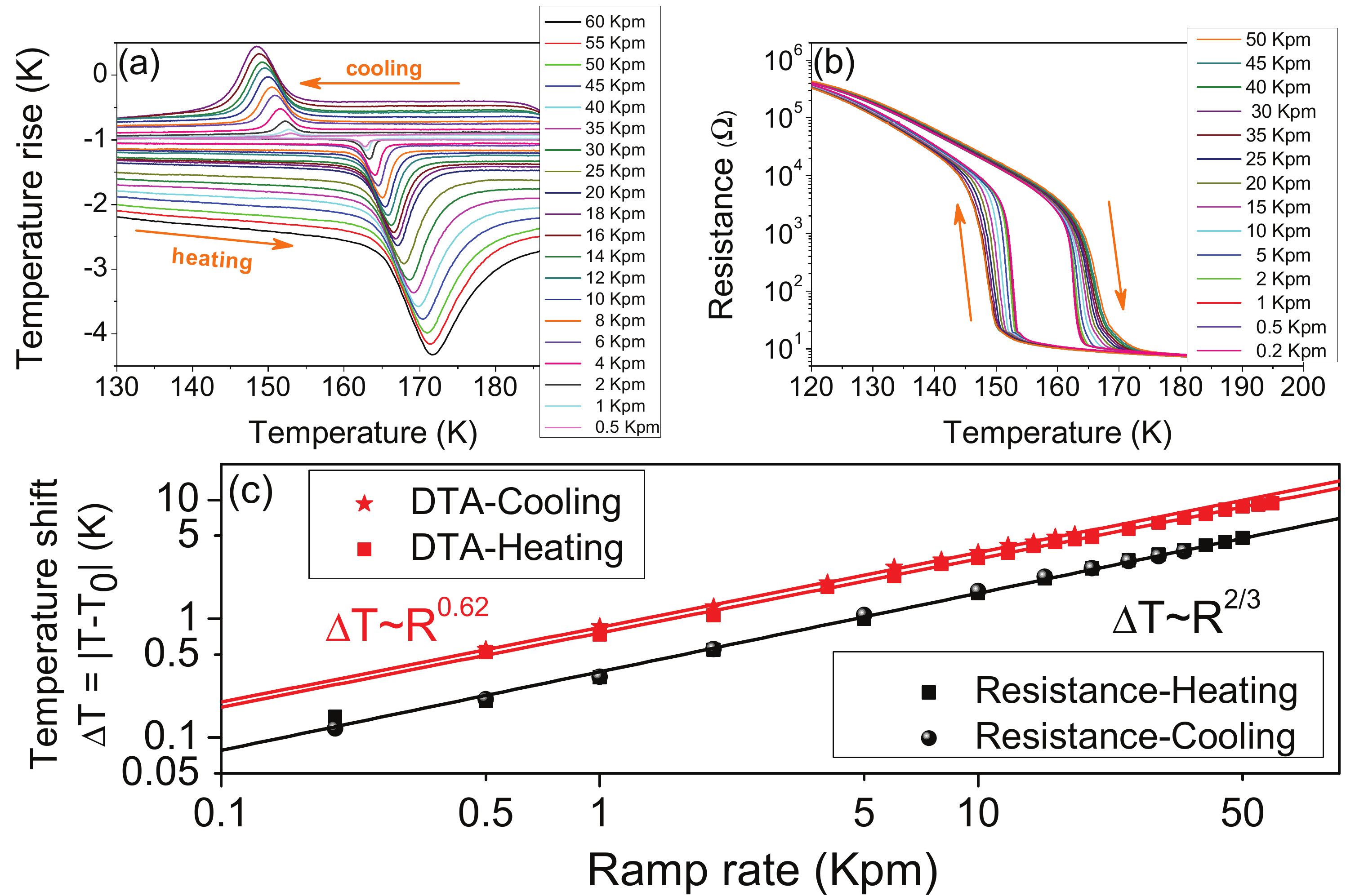}
\caption{(a) DTA signal and (b) the sample resistance as a function of temperature for different linear ramp rates. (c) Shift in the transition temperature with temperature scanning rate $\Delta T (R)$. The temperature shifts inferred from the DTA extrema are best fitted to $\Upsilon=0.62$ \cite{SupplementaryMaterial}. The transport data also approximately obey  $\Delta T (R)\propto R^{2/3}$. }
\end{figure}

\noindent
{\em Dynamic hysteresis.---}Figures 2(a) and 2(b) show the DTA and the resistance data for linear temperature ramp rates, between $0.2$ to $60$ K/min. We observe a systematic delay in the onset temperature that is dependent on the temperature ramp rate $R$. A  model-independent way to depict this dynamic shift is to plot the rate-dependent depths of supercooling and superheating. This is done in Fig. 2 (c) where the dynamically renormalized shifts $\Delta T (R)$ in the observed transition temperature $T^\textrm{heat}_\textrm{obs}(R)$ and $T^\textrm{cool}_\textrm{obs}(R)$ are seen to obey the scaling relationship $\Delta T (R)=|T^{i}_0-T^{i}_\textrm{obs}(R)|\propto R^\Upsilon$, $i=\textrm{heat}$ (heating) or $\textrm{cool}$ (cooling) over two decades of ramp rates. $T^\textrm{heat}_0$ and $T^\textrm{cool}_0$ were used as free parameters, varied to yield the best straight lines in the log-log graph \cite{SupplementaryMaterial}.  $T^\textrm{heat}_0\approx 162.77$ K and $T^\textrm{cool}_0\approx 153.35$ K thus correspond to the transition temperatures under quasistatic heating and cooling respectively. The fact that $T^\textrm{heat}_0$ and $T^\textrm{cool}_0$ are not known {\it a priori} make the estimation of $\Upsilon$ difficult. The Supplemental Material  \cite{SupplementaryMaterial} discusses this further. The values of $T^i_0$ which minimize the error in the straight line fits (on log-log scale) yield $\Upsilon\approx 0.62$ for both cooling and heating. Another independent estimate yields $\Upsilon= 0.62\pm0.06$ for heating and $\Upsilon= 0.64\pm 0.09$ (cooling) \cite{SupplementaryMaterial}.
It is significant that one should observe this symmetry. The above analysis was performed for the DTA data. It can be seen that the resistance data, where there is a greater ambiguity in extracting the actual transition temperatures, also nevertheless suggest that $\Upsilon\approx 2/3$.

To understand these observations, note that in the MF picture, the order parameter $\phi$ would evolve by the same equation that is used for {\em critical dynamics}. For nonconserved $\phi$, this is the dissipative time-dependent Landau (TDL) equation or model A \cite{chaikin-lubensky, zhong_prl2005}
\begin{equation}
{\partial \phi\over \partial t}=-\lambda{\delta \over \delta  \phi}f(\phi, T) + \zeta(t).
\end{equation}
Here $f(\phi, T)$ is the free energy, and $\lambda$ is a kinetic parameter. The stochastic force $\zeta(t)$ is zero under MF approximation. As a consequence of the above dynamics, critical-like slowing down would be observed around the transition if (and only if) the system approaches a genuine bifurcation point where the dynamic susceptibility is singular. Under the sweep of field or temperature with time, a systematic delay in the onset of phase switching is predicted with a definite scaling form. The change in the area $A(R)$ of the hysteresis loop (or, equivalently, the shift in the transition point) must dynamically scale with $R$, the rate of change of field $H$ or temperature $T$, as a power law \cite{zhong_prl2005, zhong_arxiv2015, jung, rao, krapivsky, zhang_Jphys1, Cold_atom}
\begin{equation}
A(R)=A_0 + a R^\Upsilon,
\end{equation}
where $A_0$ is the area of the quasistatic hysteresis loop. Under the deterministic evolution demanded by the MF theory, the instabilities are the spinodals,  $T^\textrm{heat}_0$ and $T^\textrm{cool}_0$ determined above. Remarkably, numerical calculations of the different (spatially averaged) free energies describing field- or temperature-induced APT all yield $\Upsilon=2/3$  \cite{jung, zhang_Jphys1, Luse-Zangwill, krapivsky, SupplementaryMaterial, footnote_otherExponents}, rather close to what we have experimentally observed.

This universality has been justified by dynamic scaling arguments \cite{zhong_prl2005}. $\Upsilon=2/3$ can indeed be recovered under the conditions of the linear ramp of the field \cite{zhong_arxiv2015} or temperature \cite{zhong_arxiv2015} if the other critical exponents are chosen to be those belonging to Fisher's $\phi^3$ theory with imaginary coupling that describes the Yang-Lee-edge singularity \cite{fisher_Yang-Lee}. This is reasonable because of the known equivalence within the MF Ising model of the Yang-Lee edge (imaginary fields, $T>T_c$) with the spinodal (real field, $T<T_c$) through analytic continuation \cite{Stephanov}.

\begin{figure}[!t]
\begin{center}
\includegraphics[clip,width=9cm]{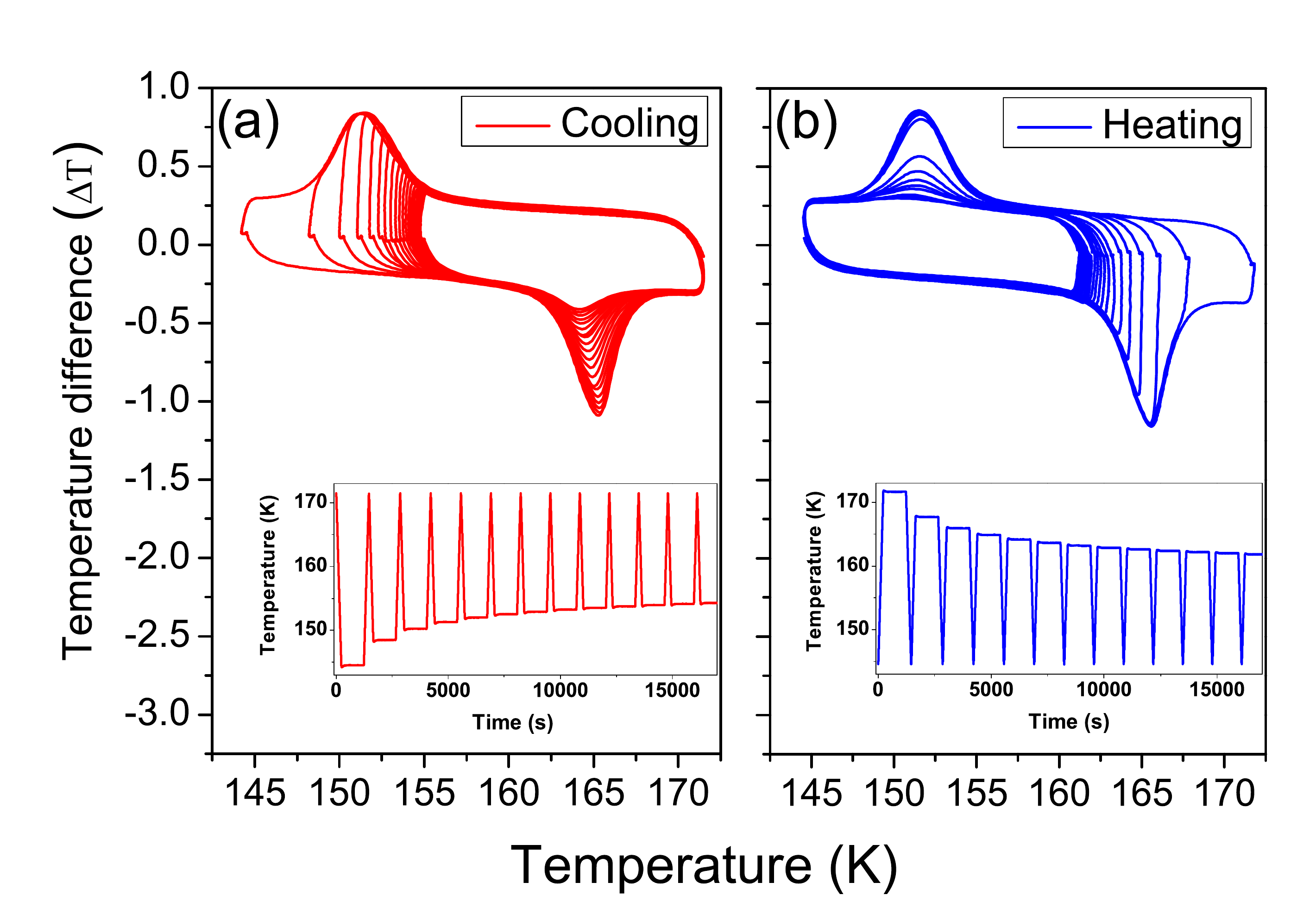}
\caption {Reversal curves for (a) cooling and (b) heating measured in DTA to map out the limits of stability and the width of the spinodal region and directly estimate the change in the order parameter across the transition. (Insets) Corresponding time dependence of temperature. (a) Starting with the initial temperature of $171.5$ K, the sample is cooled to a set temperature $T^s_i$, $i=1$. On reaching this set temperature, the system stays there for $1000$ s to reach the equilibrium value. The system is heated back to 170 K and the process is repeated with a slightly higher set temperature, such that $T^s_2>T^s_1$. The area of the latent heat peak at $\sim 165$ K is a measure of the amount of material transformed, the order parameter at $T^s_i$. This allows us map out the cooling spinodal region. (b) Shows the same idea implemented for the heating spinodal. The order parameter extracted from these measurements is shown in Fig. 4.}  \label{Fig:Fig3}
\end{center}
\end{figure}

\noindent
{\em Free energy and order parameter.---}Because of the interplay of lattice, spin, and orbital degrees of freedom that gives rise to three simultaneous transitions in V$_2$O$_3$, the temperature-induced APT is more complicated than the Ising-like transition at the Mott critical point \cite{CriticalThermodynamicsMott}. A phenomenological extension to the Ising model that will make it a temperature-driven APT and also, albeit in a rather simplistic way, capture the accompanying structural transition is the compressible Ising model \cite{domb, salinas, SupplementaryMaterial}. Here the lattice compressibility is coupled to the spin via the exchange coupling constant $J$ of the Ising model \cite{SupplementaryMaterial}. In the mean field approximation, the resulting dimensionless free energy per spin can be written as \cite{SupplementaryMaterial}
\begin{equation}
f={T\over 2T_c}[(1+\phi)\ln(1+\phi)+(1-\phi)\ln(1-\phi)]-{\phi^2\over 2 } - \xi\phi^4
\end{equation}
The scalar nonconserved order parameter $\phi$ (the average ``magnetization'' per site) is identified with the fraction of the insulating phase. $|\phi|< 1$ at any nonzero temperature and $T_c$ is the critical temperature. For ${T\over 12T_c}-\xi<0$, one would observe an APT during thermal cycling.

Figure 3 shows a method to experimentally estimate $\phi$  at the given temperature using DTA. $\phi$  is taken to be proportional to the integrated area around the DTA dip (peak) for the cooling (heating) curves in Figs. 3(a) and 3(b) respectively, as a function of the temperature of approach and is plotted in Fig. 4. A $1000$ s wait at the temperature of approach ensures quasistatic conditions. Also in Fig. 4, $\phi$ is independently estimated from the resistance data using McLachlan's effective medium theory \cite{McLachlan} to approximately handle the percolative nature of the transport \cite{SupplementaryMaterial}.

Remarkably the two free parameters of the model, $\xi$ and $T_c$, are already fixed by the experimentally inferred spinodal temperatures. $T_c\approx 153.5$ K and the value of $\xi$ is numerically determined to be $0.154$ from the value of the other spinodal temperature to be $162.5$ K. Thus to describe the dynamics [Eq. (1)], the remaining free parameter $\lambda$ is also fixed by fitting any one of the dynamic hysteresis curves;  $\lambda=3.5$ s$^{-1}$ was obtained by fitting the curve for the inferred order parameter (fraction of the insulator phase) by evolving Eq. (1) for heating with a linear temperature  ramp at the rate of $50$ K/min. In the simulation discussed in the Supplemental Material \cite{SupplementaryMaterial}, a hysteresis scaling exponent of 2/3 is observed, as is expected from generic arguments given above. The results for the inferred order parameter from the compressible Ising model are also shown in Fig. 4. The transition was given a very small but finite width by assuming that the sample is an inhomogeneous ensemble with a Gaussian distribution of $T_c=153.8$ K with a standard deviation of $0.18$ K.

\begin{figure}[!t]
\includegraphics[clip,width=8cm]{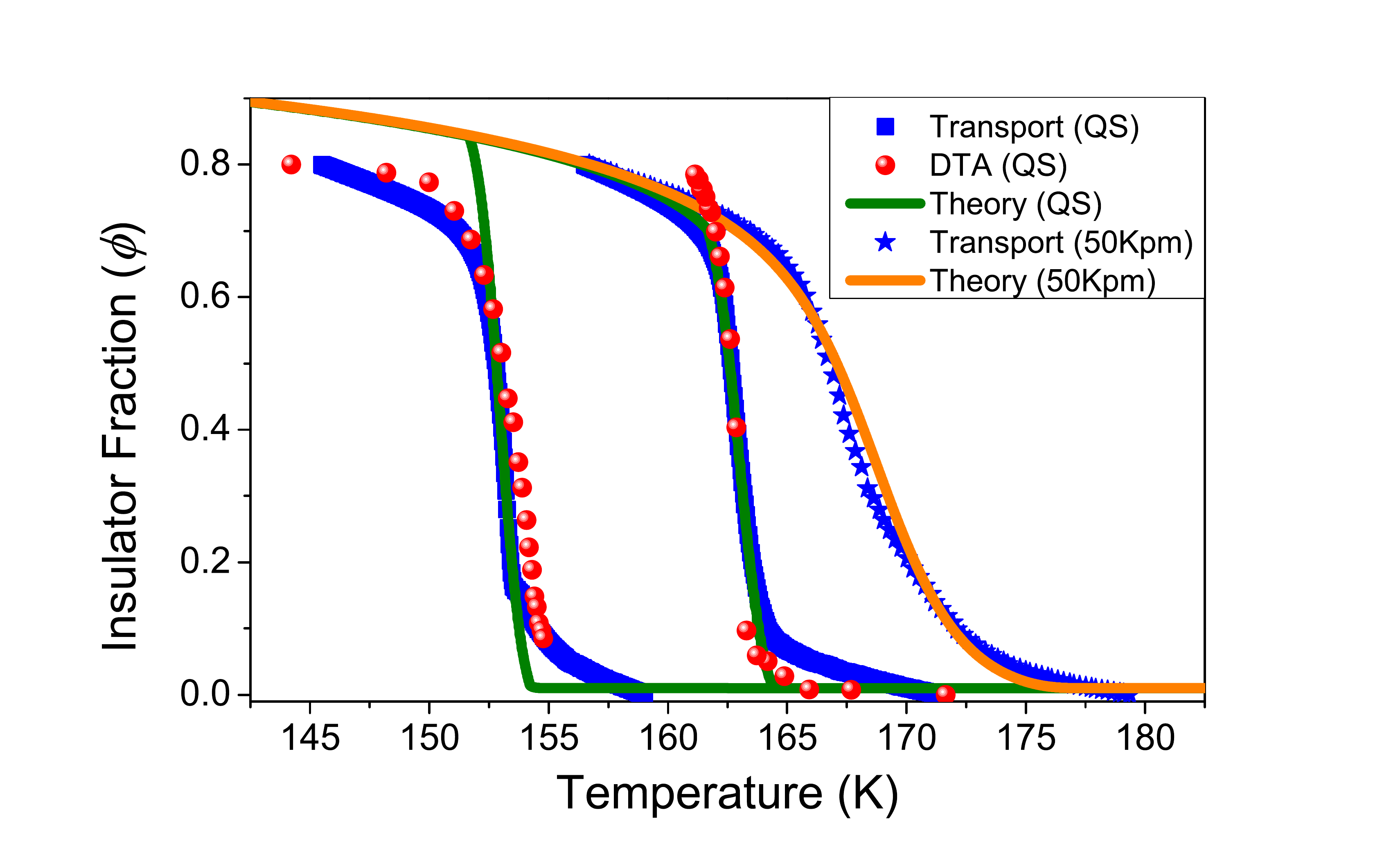}
\caption{($\circ$) The quasistatic (QS) order parameter $\phi$ vs temperature extracted from DTA  using data in Fig. 3. ($\Box, \ast$) $\phi$ inferred from resistance measurements.
(---) Compressible Ising model.}  \label{Fig:Fig4}
\end{figure}

\noindent
{\em Phase ordering in quench-and-hold experiments.---}Starting from the initial temperature of $100$ K ($240$ K), the experimental contour plots in Fig. 5 are obtained by rapidly (at the rate of $50$ K/min) heating (cooling) the sample to different target temperatures  $T_w$ slightly above (below) the quasistatic transitions temperatures \cite{SupplementaryMaterial}. Once $T_w$ was reached, the temperature was kept constant and the time evolution of the resistance at different $T_w$ is plotted in terms of the insulator fraction \cite{McLachlan, SupplementaryMaterial}. Figure 5 demonstrates that (in a qualitative sense) the phase transformation proceeds symmetrically, with similar timescales. Given that the metastable phase is bounded also on the low-temperature side makes the physics of arrested kinetics of the  metastable phase qualitatively different from that observed in glasses.

The corresponding calculation [using Eqs. (1) and (3)] for the noise-free evolution of the order parameter after shock heating is also shown in Fig. 5. With $T_c$, $\xi$, and $\lambda$ already fixed, the entire contour plot has no free parameter. The qualitative match with the highly constrained  MF calculation supports the picture of phase transformation occurring via barrier-free continuous ordering. Calculation for the cooling quench, due to its sensitivity on the initial conditions, is not discussed.
\begin{figure}[!t]
\includegraphics[clip,width=8cm]{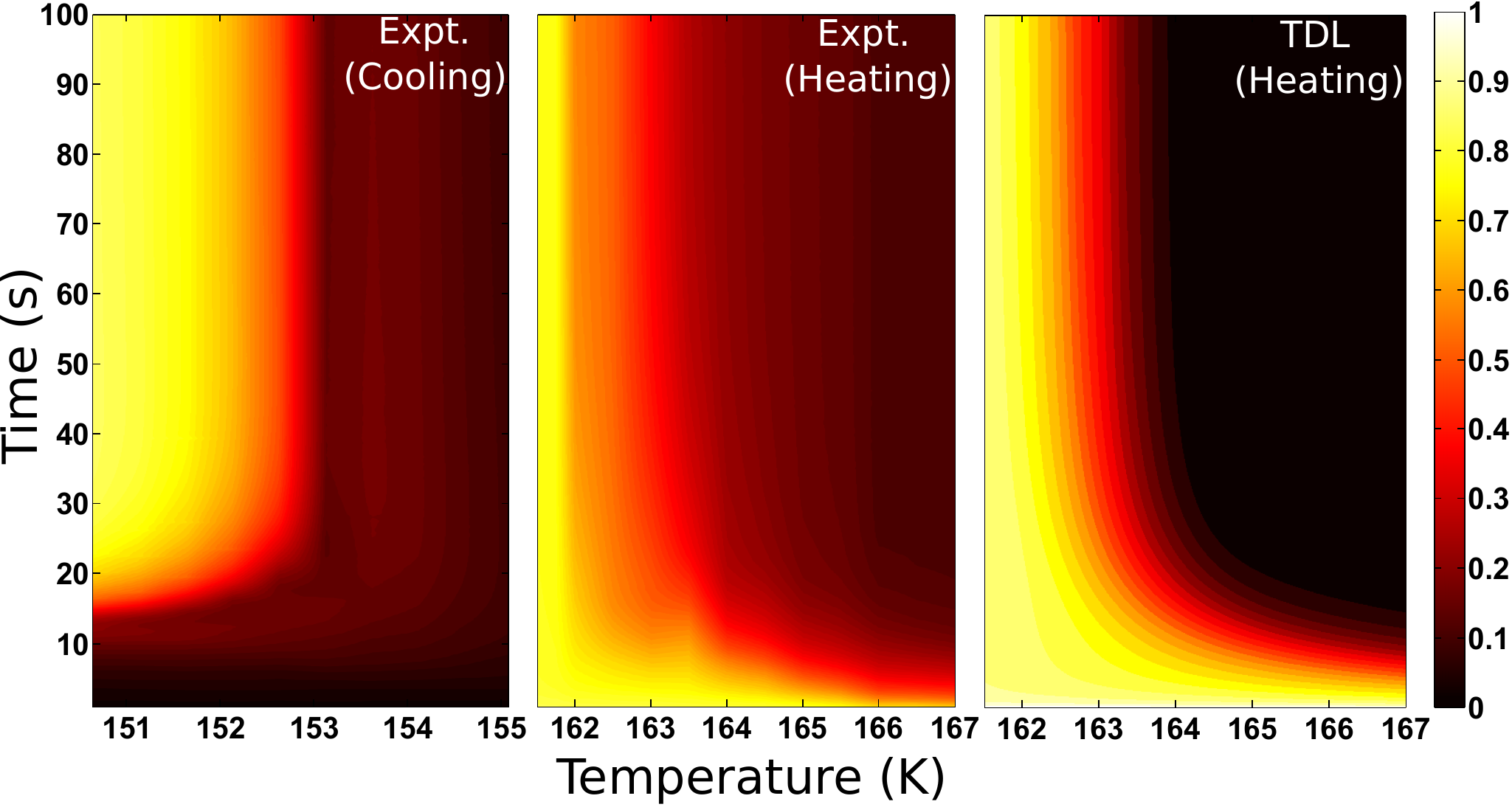}
\caption{Critical-like quench-and-hold dynamics of the order parameter (insulator fraction) around the cooling and heating spinodals. The data show the first $100$ s of the time evolution after the target temperature $T_w$ (denoted by the abscissa) has been reached \cite{SupplementaryMaterial}. The calculation involves evolving the compressive Ising model using model A [Eq. (1)] with no free parameters \cite{SupplementaryMaterial}.} \label{Fig:Fig5}
\end{figure}

\noindent
{\em Conclusions.---}The unreasonable efficacy of the MF theory in capturing the essence of the transformation in V$_2$O$_3$ gives credence to the idea that---and this is the key result of our work---spinodal-like instabilities can be present in a real material exhibiting a finite-temperature abrupt phase transition. Scaling of dynamic hysteresis with the observed exponent and barrier-free phase ordering are both manifestations of these instabilities.

At least for phase transformation under such deep supersaturation, recent work on very different aspects of the problem suggests that fluctuations may not fundamentally affect these qualitative aspects.
Within the Ising model in thermal {\em equilibrium} (and $T<T_c$), the MF spinodal corresponds to the two values of magnetic field demarcating the limit of metastability. Rigorous mathematical analysis shows that the effect of fluctuations (or equivalently making the range of interactions finite) is simply to rotate this spinodal magnetic field in the complex plane, giving it a nonzero imaginary value \cite{Stephanov, Gulbahce-Gould-Klein}. Thus, in analogy with the Yang-Lee argument that the complex zeros of the partition function only touch the real axis in the thermodynamic limit, fluctuations essentially mimic finite-size effect  \cite{Gulbahce-Gould-Klein}. Remnants of this singularity should be discernable in the broadened transition if the range of the potential is large enough, as it might be for V$_2$O$_3$ due to the deep supersaturation. Furthermore, for a dynamically changing control parameter, the system may get too sluggish in the vicinity of the transition (because of the critical-like slowing down) to turn on these fluctuations before the transition has occurred. Numerical solutions of model A [Eq. (1)] now also including the stochastic term ($\zeta\neq 0$) do indeed show that fluctuations only slightly change the value of the exponent \cite{zhong_prl2005, berglund} describing this dynamic overshoot, still not far from our observations.

While these arguments make the experimental observations plausible, it is emphasized that the metastable phase is properly only to be defined in a dynamical sense. Dynamically emerging spinodal-like singularities have been seen in simulations of Lennard-Jones fluids \cite{trudu, Maibaum, santra-bagchi}, binary alloys \cite{Razumov}, elementary models \cite{Pelissetto, zhong_pre2017}, and perhaps even in some other experiments \cite{Collins-teh}. Thus, while the precise nature of the kinetic spinodals is yet to be determined, their existence in specific contexts seems real enough. These instabilities have been variously interpreted---for example, as the boundary between the regions of the homogeneous and heterogenous nucleation \cite{Razumov}.

Such strongly hysteretic ``zeroth-order'' transitions \cite{footnote zeroth-order} thus form a new class of transitions that may potentially be observed in many other systems including similar oxides undergoing metal-insulator transition \cite{Alsaqqa, liu}, manganites \cite{levy}, intermetallic shape-memory alloys and magnetocaloric materials \cite{Liu_Heusler}. A better understanding of the phase transformation kinetics in such systems should help chart the uncertain territory of metastable states in the language of critical phenomena.

It is a pleasure to thank Subodh R. Shenoy, H. R. Krishnamurthy, and especially Fan Zhong for helpful comments and suggestions.

\newpage
\begin{center}
{\Large
{\bf Supplemental Material}
}
\end{center}
\tableofcontents

\section{V$_2$O$_3$ samples}
For this study we have used polycrystalline V$_2$O$_3$ powder, procured form Sigma-Aldrich Corporation. This was pelletized for thermal and transport measurements. That the material was single phase was verified by X-ray diffraction (XRD) [Fig. 6]. The SEM micrograph [Fig. 7] shows that (while there is a dispersion in the grain size), most grains are well above $\sim 500$ nm and the sample can be considered to be bulk material.

\begin{figure}[h!]\label{figure1}
\center
\includegraphics[scale=0.3]{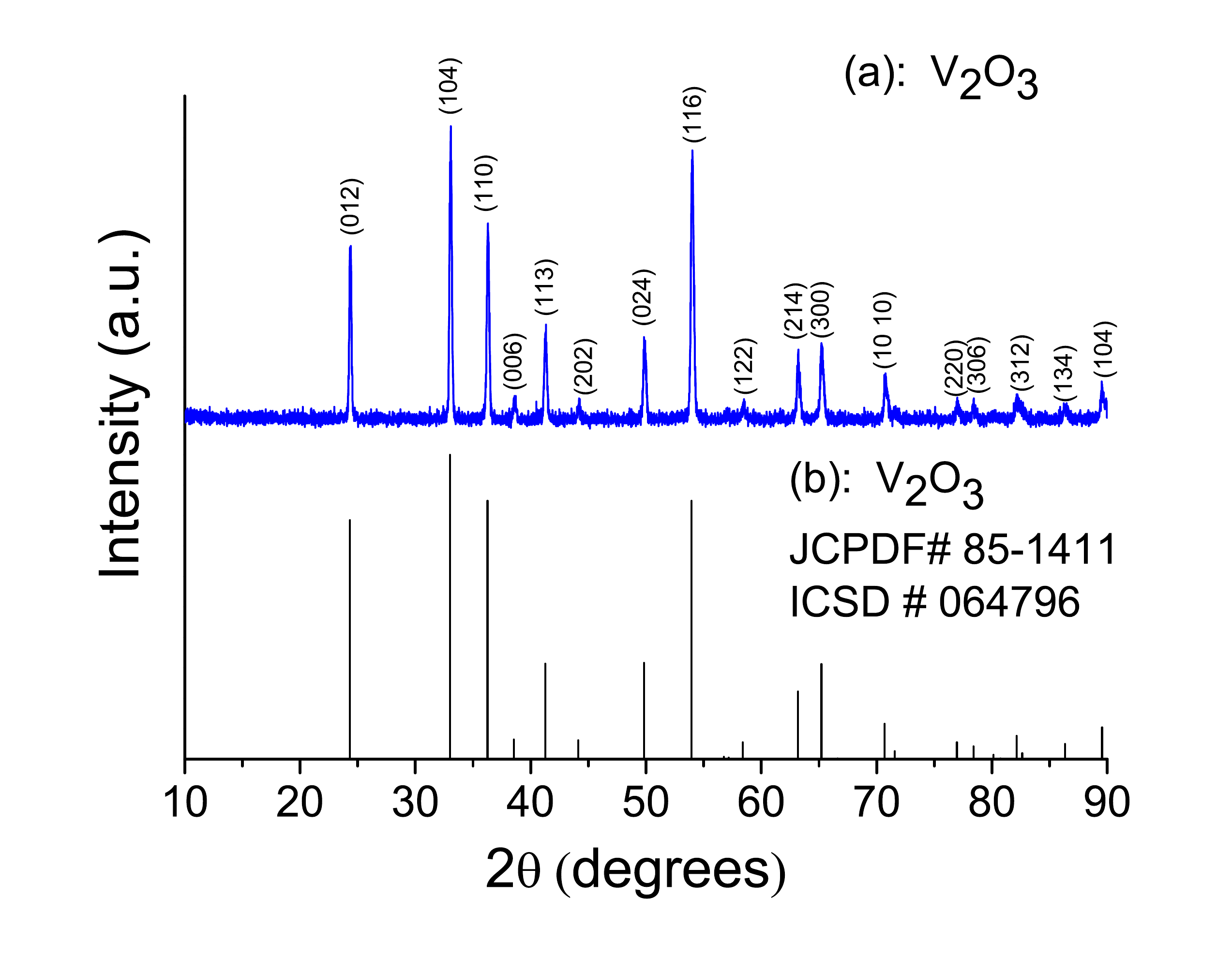}
\caption{(a) Room temperature XRD of V$_2$O$_3$ for Cu $K_\alpha$ ($\lambda = 1.54178$ \AA) radiation, (b) standard V$_2$O$_3$ XRD taken from
 the Joint Committee on Powder Diffraction Standards (JCPDS) data corresponding to the rhombohedral form (room temperature). }
\label{Xrd}
\center
\end{figure}

\begin{figure}[b!]
\center
\includegraphics[scale=0.24]{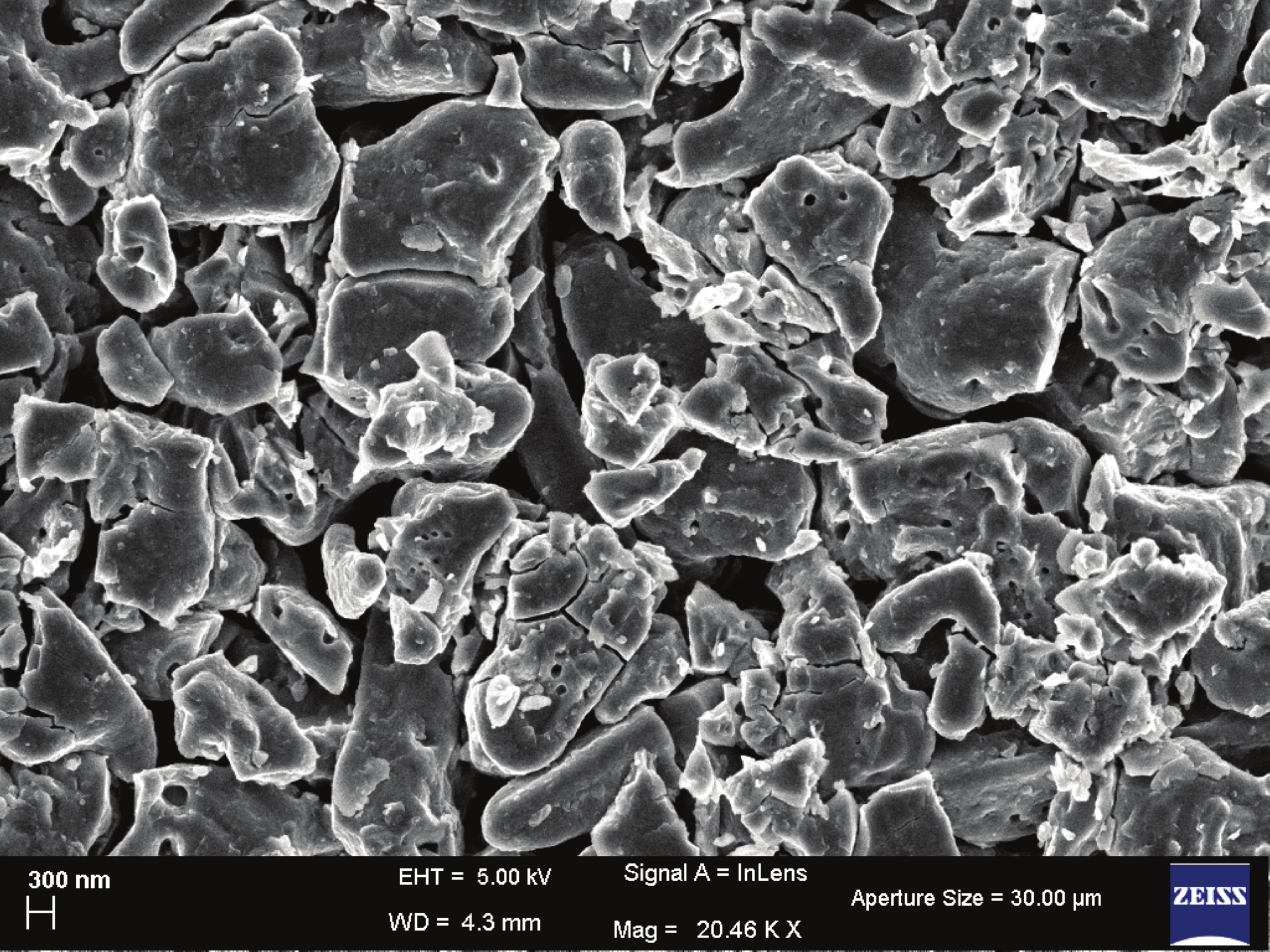} \label{figure 2}
\caption{Scanning electron micrograph of the V$_2$O$_3$ powder.}
\label{Xrd}
\center
\end{figure}

Note that polycrystallinity of the samples should not affect the differential thermal analysis measurements (DTA), where it is more important to have high-purity strain-free single-phase material. These attributes were confirmed by the sharp endo- and exothermic DTA peaks [Fig. 1(a) (main text)].

On the other hand, the observed resistance in the high temperature metallic phase is likely to be dominated by the grain-boundary resistance. Indeed, while the resistivity of the insulating phase matched that of pure single crystals, it was two-three orders of magnitude higher in the metallic phase. As the observed change in the resistance at the transition was still $2-3$ orders of magnitude, having pelletized polycrystals (which in the context of this study are preferable to thin films because of absence of strain and a sufficient volume required for DTA measurements) does not affect any of the conclusions of this work. Most importantly, the observed transition temperatures during (quasistatic) heating and cooling cycles are in excellent agreement with those previously reported on single crystals.

\section{Experimental details: transport}
\subsection{Transport experiments}The experiments were done in a liquid nitrogen cooled variable temperature insert. As the sample resistance varied by more than five orders of magnitude between room temperature and $77$ K, special care was taken to make the measurements reliable. Resistance measurements were performed by exciting the sample with an alternating voltage source (frequency $31$ Hz) and using $two$ lockin amplifiers. A 1 k$\Omega$ resistance was kept in series with the sample and the sample current and sample voltages were measured by measuring the voltages across the standard resistance and the sample respectively. The excitation voltage was varied such the voltage drop across the sample was $\sim 1-10$ mV. The sample voltage was measured with a voltage preamplifier with input impedance of 100 M$\Omega$.

When starting with the virgin sample, the sample resistance (even at room temperature, far away from the metastable hysteretic region) was initially history dependent. But after a few tens of thermal cycles, the resistance stabilized to a reproducible history-independent value at 300 K. The measurements reported here were all performed on such `trained' samples. This phenomenon is well known for materials undergoing martensitic transitions and is likely to be due to the formation of microcracks \cite{footnote-microcracks}.

\subsection{Reliability of the dynamic hysteresis data}
 In order to rule out any cryostat-related artifacts in observed temperature scanning rate dependent shifts in the transition temperature, we have measured the resistance upon both heating and cooling of a BaFe$_2$As$_2$ sample for similar temperature ramp rates. BaFe$_2$As$_2$, the parent compound for pnictide superconductors, has a non-hysteretic spin-density wave transition $T_c\sim 135$ K where the sample resistance shows an abrupt change in slope \cite{Kim-Kim}.  Fig. 8 shows that there is no difference in the transition temperature in the data taken during heating and cooling. Nor is there any rate-dependent shift in the transition temperature, as temperature ramp rate is varied between $5-40$  K/min.
\begin{figure}[h!]\label{BaFe2As2}
\center
\includegraphics[scale=0.3]{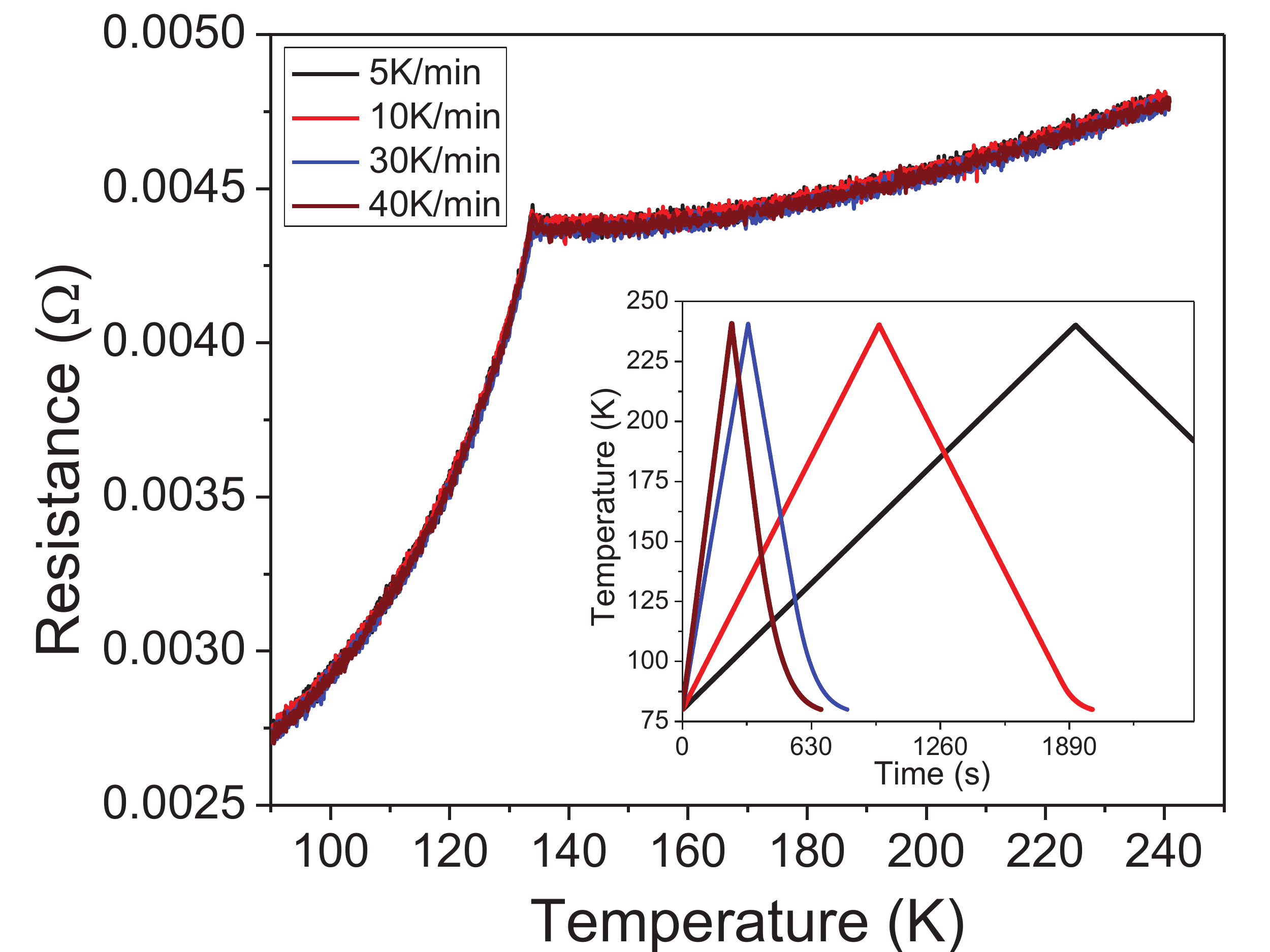}
\caption{The spin-density wave transition temperature in a BaFe$_2$As$_2$ sample is independent of the temperature sweep rate. (inset) The time dependence of the temperature sweep at the rates of $5,\, 10,\, 30,$ and $40$ K/min.}
\label{BaFe2As2}
\center
\end{figure}

\subsection{Conversion of resistance to insulator fraction}
Estimation of the insulator fraction $\phi$ from the resistance data requires the use of a percolation model. We have used  McLachlan's  general effective medium theory \cite{McLachlan}, which has been successfully used in previous transport studies on three dimensional metal-insulator mixtures \cite{PhysRevLett.kim_2000}. The insulator fraction $f$ is given by
\begin{equation}\label{eq:2}
f\frac{(\sigma_I^{1/t} - \sigma_E^{1/t})}{(\sigma_I^{1/t} + A\sigma_E^{1/t})} + (1-f)\frac{(\sigma_M^{1/t} - \sigma_E^{1/t})}{(\sigma_M^{1/t} + A\sigma_E^{1/t})} = 0.
\end{equation}
 $\sigma_I$ and $\sigma_M$ are the conductivities of insulating and metallic phases, and $A = (1-f_c)/f_c$, $f_c$ is the insulator fraction at the percolation threshold and $t$ is the critical exponent. The value of  $f_c$ and $t$ depend on the dimension of the system \cite{PhysRevLett.kim_2000, PhysRevB.schuller_2009}; for three dimensions $f_c \approx 0.16$ and $t \approx 2$.

\section{Temperature stability during phase-ordering dynamics}
In the experiments corresponding to Fig. 5 (main text), we first quench (or shock heat) to a target temperature $(T_w)$ and observe the time dependent relaxation at this temperature by monitoring the sample resistance. Short of dunking the sample in a liquid bath, quickly settling at the given set temperature after a rapid quench is difficult to achieve experimentally. The temperature stabilization always takes a finite time and the temperature is prone to oscillations during this phase.

The quench rate was chosen to be $50$ K/min, the same for all the data plotted in Fig. 5 (main text). This was the largest rate for which linear temperature ramp could be accomplished in our cryostat for both cooling and heating. Fig. 9 shows the actual time dependence of the sample temperature for different $T_w$ used to make the contour plot [Fig. 5 (main text)]. The quality of control (how well and how quickly the temperature stabilized on reaching the target wait-temperature $T_w$ after quench), though not still quite perfect, required considerable effort in determining the best `proportional-integral-derivative' (PID) control gain values of the temperature controller, as well as, in finding the  optimal exchange gas pressure in the cryostat

The temperature fluctuation for the shock-heating data is about $\pm 0.2$ K. Despite this degree of control, even such small fluctuations were sufficient to slightly blur the contour plot [Fig. 5 (main text)] around $163$ K, where the change in resistance with time was maximum. As the stability for the cooling quench experiments was worse $(\sim 1\textrm{K})$, so here it was ensured that at least the temperature did not oscillate. For the cooling quench experiments, the lowest temperature reached was taken to be $T_w$.

\begin{figure}[h!]\label{Fig_QW_Transport}
\center
\includegraphics[scale=0.4]{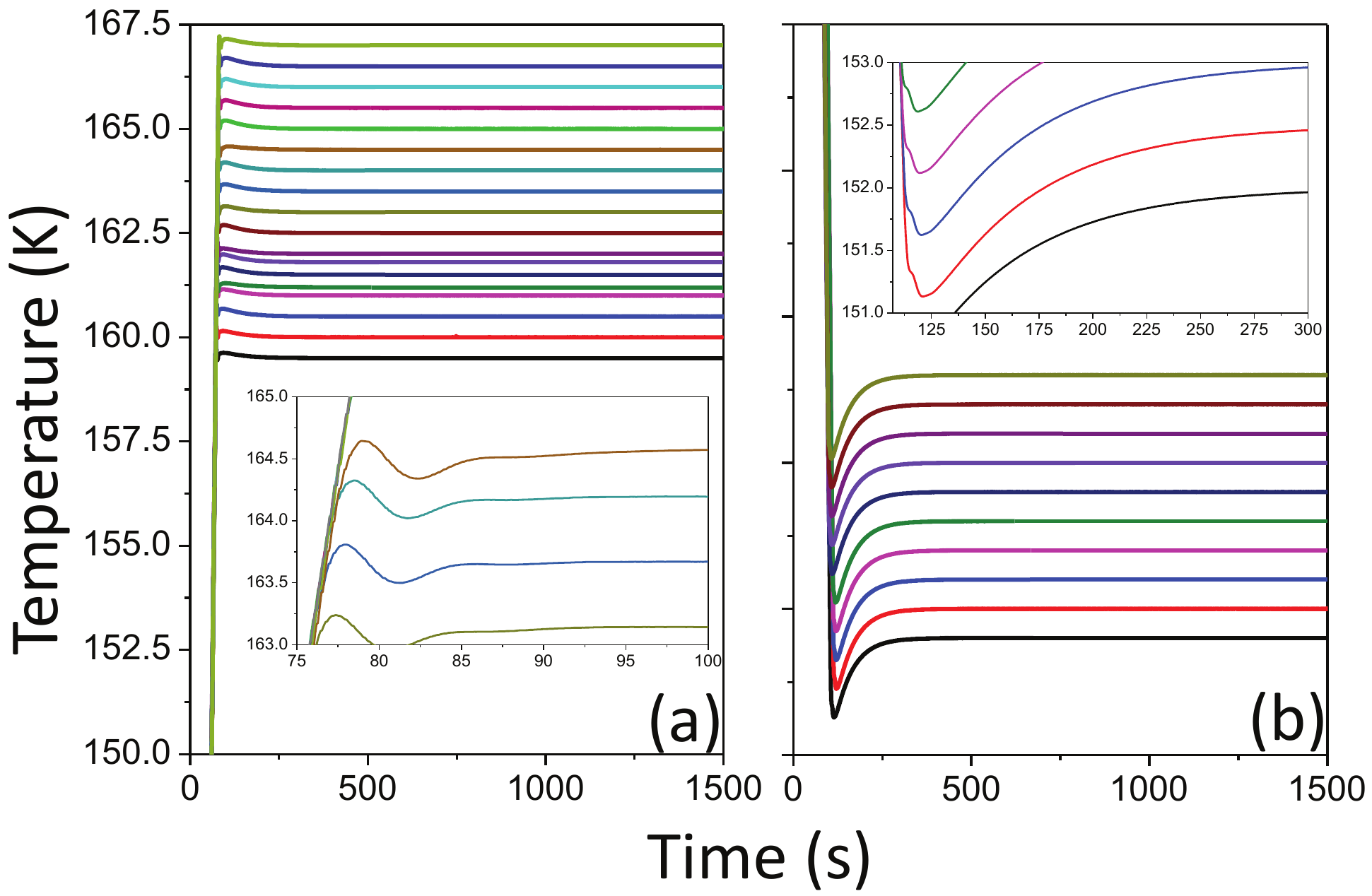}
\caption{Temperature stability of `quench-and-hold' experiments for (a) heating (b) cooling. Note that the data for heating is much better. This is a lucky coincidence because experiment-theory comparison could also only be made for the heating data. }
\label{QAW_transport}
\center
\end{figure}

\section{Differential Thermal Analysis}
Fig. 10 (a) shows the schematic of our homemade calorimeter placed inside an Oxford Instruments cold-finger type liquid nitrogen cryostat. Two identical substrates (heat capacity $C_{sub}\approx$ 2.75 mJ/K) consisting of thin films of platinum (standard Pt-100 resistance thermometers) are used to detect temperature \cite{Nagapriya, Schilling}. The two substrates are connected to a temperature-controlled copper thermal bath (heat capacity $C_{bath}\approx$ 31.74 J/K) through a poor thermal link (a glass cover slip) of thermal resistance ($R_{th}$).

\begin{figure}[h!]\label{Fig_DTA_Schematic}
\center
\includegraphics[scale=0.32]{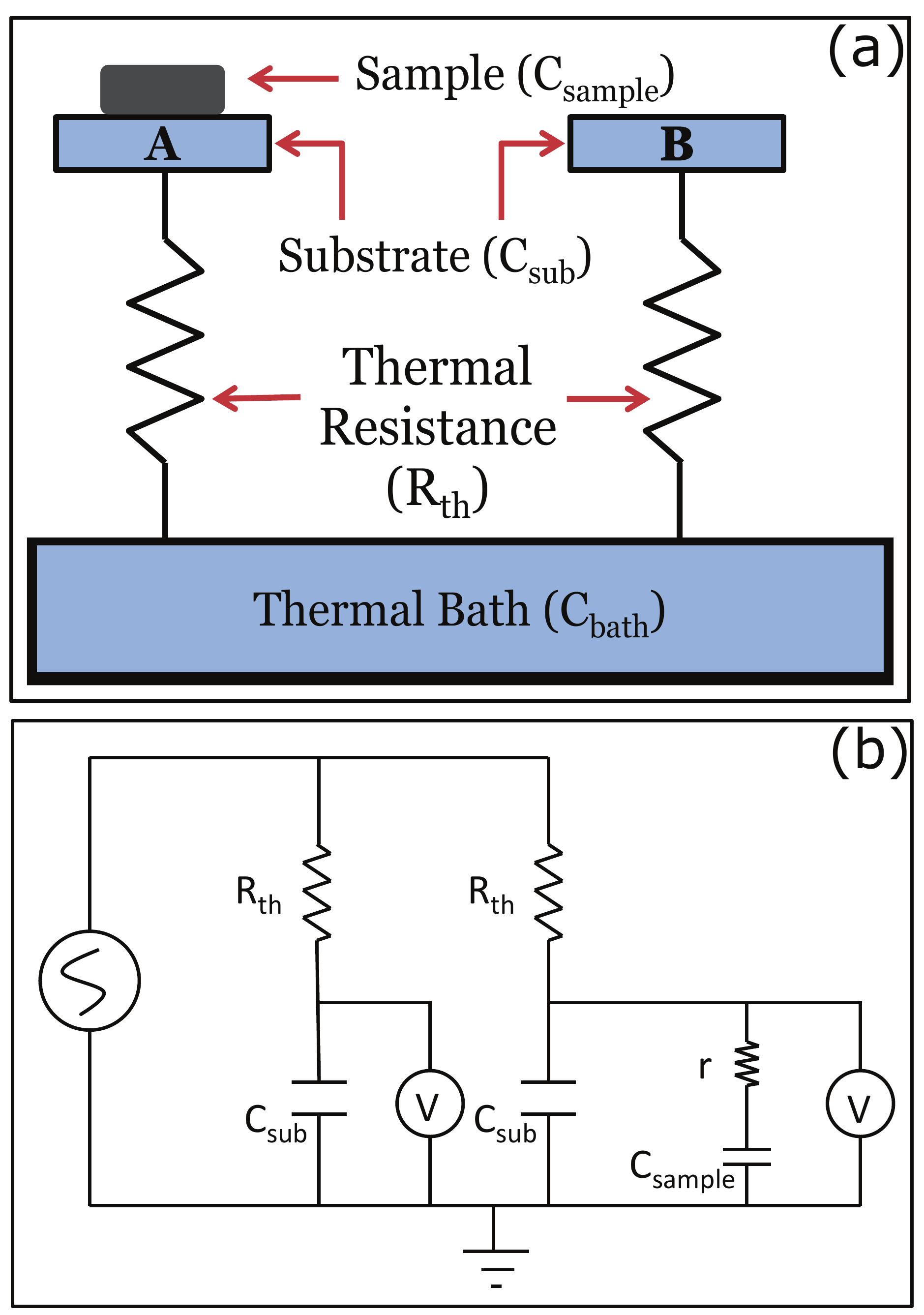}
\caption{(a) Schematic of the calorimeter, (b) Equivalent electrical circuit of calorimeter based on the mathematical analogy between the Biot-Fourier law and Ohm's law.}
\label{Figure_2}
\center
\end{figure}

Based on the thermal-to-electrical analogy coming from the similar structure of Biot-Fourier and Ohm's laws, there is a straight-forward  one-to-one mapping between the components of the thermal circuit and an equivalent electrical circuit. Fig. 10 (b) shows this equivalent electrical circuit of the calorimeter [Fig. 10 (a)]. The heat capacity, temperature, heat, heat flux, and thermal resistance in the thermal circuit correspond respectively to the capacitance, voltage, charge, current and resistance in  the equivalent electrical circuit.

Due to the much larger heat capacity of the temperature-controlled copper block (in comparison to the sample and the substrate heat capacities, $C_{sample}$ and $C_{sub}$) it can can be thought of as a thermal bath (or a constant voltage source in the electrical analogy).  The time constant (relaxation time) for the sample-substrate system to reach the temperature of the bath was determined to be $\tau = R_{th}C_{sub}\approx 3.2$ s.

In Fig. 2 (a) [main text], there is a systematic enhancement in the areas of the latent heat peaks with the temperature ramp rate. It is naturally of interest to explore if some conclusions about the difference in the latent heat as a function of ramp rate can be made from this observation. Unfortunately, the relationship between the observed  ramp-rate dependent area of latent heat peak and the actual latent heat released/absorbed in the experiment is not quite straightforward. Fig. 11 shows the simulated result of the equivalent circuit where the temperature is linearly swept at different rates and the same amount of heat is released/absorbed at the transition temperature as a $\delta-$function heat pulse. Since the areas of the simulated peaks are also observed to be dependent on the temperature ramp rates, we conclude that it is nontrivial to unambiguously estimate  the magnitude of the latent heat from the area of the peak. The sensitivity of the area of the latent heat peak on the temperature scanning rate also comes from the additional fact that, experimentally, the transition is not quite abrupt and the recorded area of the latent heat peaks is diminished at low ramp rates as some heat has already escaped/entered the substrate before all the latent heat has been released/absorbed.
Hence, throughout this work, we have only focussed on the value of the transition temperature when comparing DTA experiments done with different ramp rates.
\begin{figure}[h!]\label{Fig_DTA_Simulation}
\center
\includegraphics[scale=0.4]{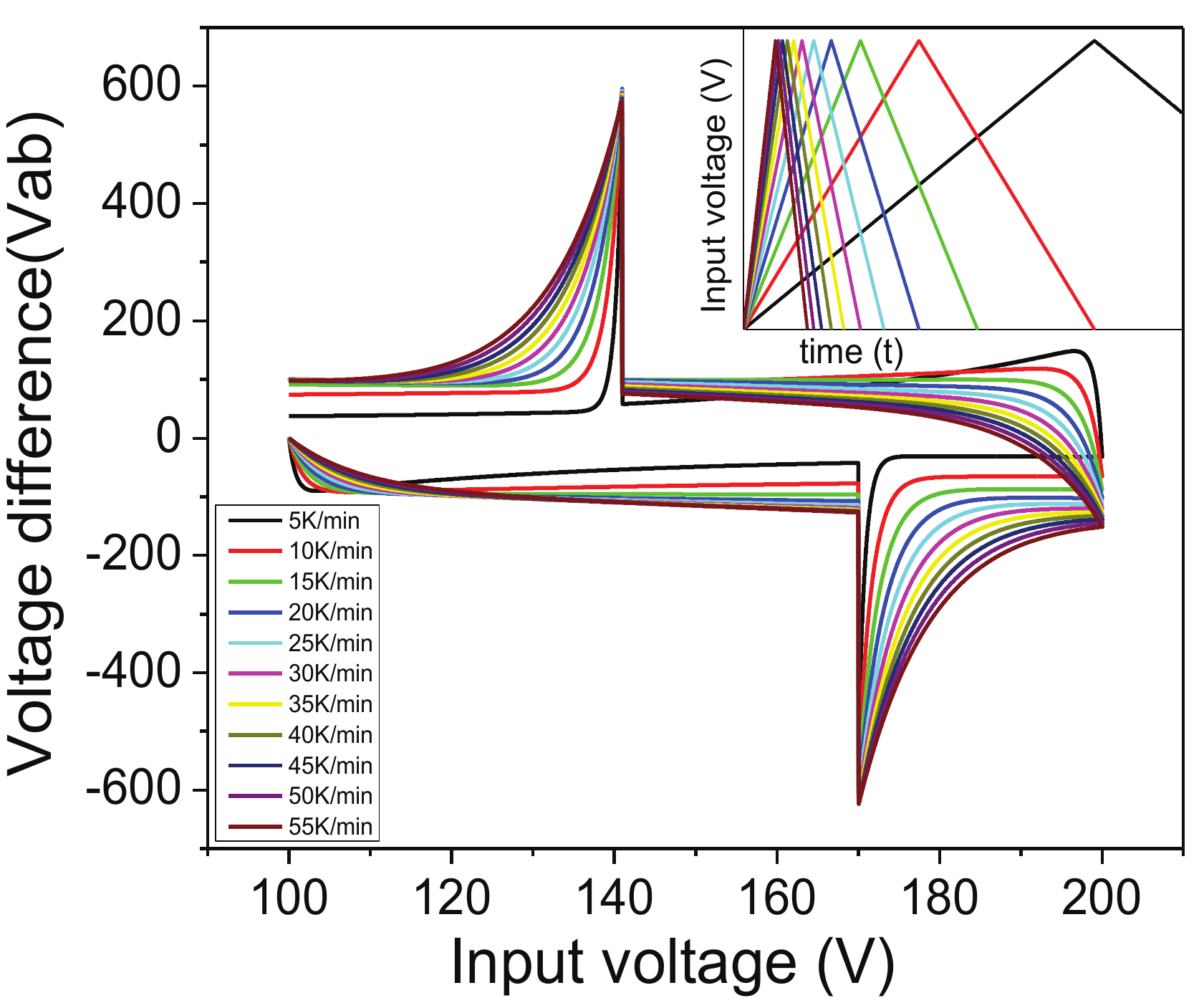}
\caption{Simulation of the DTA signal using the equivalent electrical circuit [Fig. 10 (b)] for different temperature ramp-rates. While the amount of latent heat released/absorbed for each case is the same, the simulation of the area of the measured DTA peaks is different for each ramp rate. Moreover there is no simple relationship between the temperature ramp rate and the DTA peak areas. (inset) Time vs temperature profiles used in the simulation.}
\label{Theoretical_Model}
\center
\end{figure}

\section{Experimental setup: DTA}
\subsection{Latent heat measurement of \boldmath V$_2$O$_3$}
The actual electric circuit for the measurement electronics is shown in Fig. 12.
\begin{figure}[h!]\label{Fig_Measurement_setup}
\center
\includegraphics[scale=0.3]{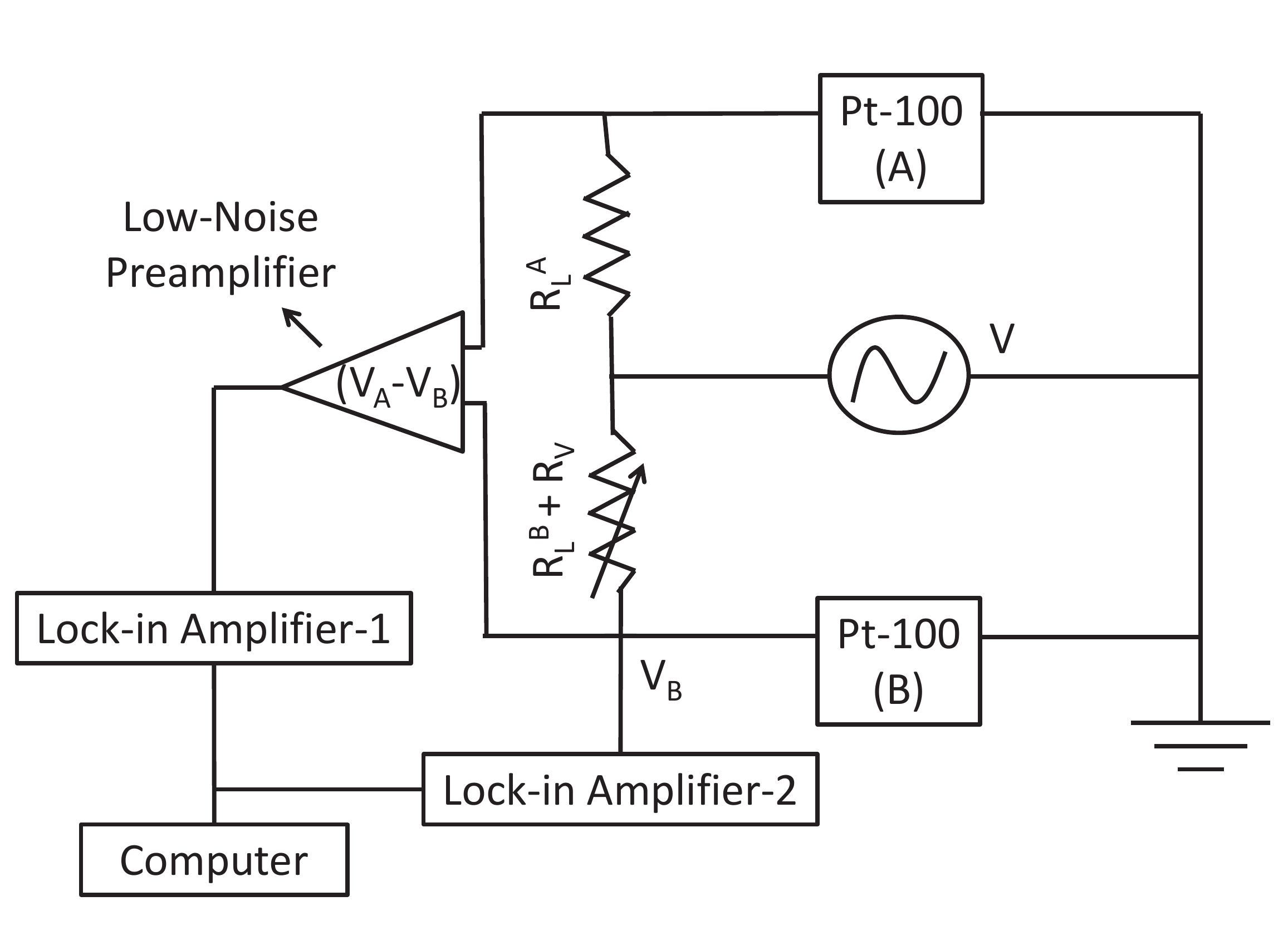}
\caption{Schematic of the electronic circuit in the latent heat measurement setup. The difference in the temperatures between the two Pt-film resistance thermometers, $A$ and $B$, with $A$ also acting as the substrate for the calorimeter is measured as the difference voltage in the bridge circuit. The resistance of reference substrate $B$ is also separately monitored to infer the absolute temperature.}
\label{Measurement_setup}
\center
\end{figure}

The calorimeter in the set up consists of two nearly identical calibrated Pt-film resistance thermometers, the substrates $A$ and $B$ of the previous figure. One them is mounted with the sample and the other one provides the reference. Sinusoidal voltage (V) of (frequency $\sim 1$ kHz) is passed (from lock-in amplifier) through load resistor $R_L^A$ and $R_L^B$ ($R_L^B \approx R_L^A$) to these substrates A and B respectively. A small variable resistor ($R_V$) is connected with $R_L^B$ to establish a bridge arrangement which enhances the sensitivity of the experiment. The voltage across the reference substrate is measured using a lock-in amplifier while the voltage-difference  between the sample substrate and the reference substrate is measured using another lock-in amplifier after passing through a low-noise voltage preamplifier. The temperature change in the calorimeter leads to a change in the resistance of the Pt-film. The resistance of the Pt-film can be calculated from the measured voltage (current is constant) of the two substrates. The absorption/release of the latent heat during the abrupt phase transition leads to change in the relative temperatures $(\triangle T)$ between the substrates $A$ and $B$. These are the latent heat peaks observed in Fig. 1 (a) (main text),  Fig 2 (a) (main text) and Fig 3 (main text).

\section{Fitting the dynamic hysteresis exponent}
\subsection{Method 1: Best straight line fits on a log-log plot}
The dynamic hysteresis exponent  $\Upsilon$ was extracted from fitting the  transition temperatures $T_i$ observed for different temperature ramp rates $R$ to the following equation
\begin{equation} \label{eq1}
T_{i} = T_0 \pm aR_{i}^\Upsilon
\end{equation}
 which has three unknown parameters, $T_0$ the quasistatic transition temperature for cooling in the case of fitting the data during the cooling runs or heating when fitting the data to the heating run, $\Upsilon$, and the constant $a$. The plus sign corresponds to an increase observed during heating and negative to the ramp-rate dependent cooling experiments. Given that the measurements were done over more than two decades of $R$, the prescription to fit would be to plot $|T_{i}-T_0 |$ versus $R$ on the log-log scale. Then $\Upsilon$ would just be the slope of the straight line.
While the value of $T_0$ is known within less than a kelvin, very small changes in $T_0$ can lead to large shifts in the low ramp-rate data because of the log scale. This problem is well-known from the early days of critical phenomena research. To illustrate this point, Fig. 13 shows how small differences in the values chosen for $T_0$ yield different values of $\Upsilon$. Although one can still fit `acceptable' straight lines, it is evident that the goodness of fit is heavily compromised if the values of $T_0$ are out of an interval. So we have estimated $\Upsilon$ using a value $T_0$ that yields minimum error in the slope [Fig. 14].

We can do a further consistency check as there are rough bounds on the acceptable value of $T_0$. $T_0$ bounded on one side by the observed transition temperature under the smallest (non-zero) ramp rate. Furthermore, the magnitude of the temperature shift should be a monotonically increasing function of $R$. This implies, for example that the shift in the transition temperature in going from a ramp rate of $1$ K/min to $0.5$ K/min should be more than the shift in going from $0.5$ K/min to the quasistatic curve. Hence we independently have an estimate of the window of values for acceptable $T_0$. Fig. 14 (b) shows that the values of $T_0$ where $\Upsilon$ has minimum error are indeed acceptable, for cooling as well as heating curves. Based on Fig. 14, we have estimated $\Upsilon\approx 0.62$.

The analysis depicted in Fig. 13 and Fig. 14 is on the data from DTA measurements [Fig. 2 (a) (main text)]. We have taken the DTA measurements for the transition temperature to be more reliable as the transition can be attributed to the extrememum in the DTA signal, and we have observed relatively sharp single peaks in the DTA measurements. The resistance data in Fig. 2 (b) (main text) should only be treated as qualitative, especially because the inferred transition temperature is dependent on the resistance value used to delineate the transition. This is to an extent arbitrary and can lead to a small difference in the inferred shifts depending on the cutoff. Nevertheless, it can be seen that the slope is very close to 2/3.
\begin{figure}[h!]\label{Fig_DiffSlopes}
\center
\includegraphics[scale=0.3]{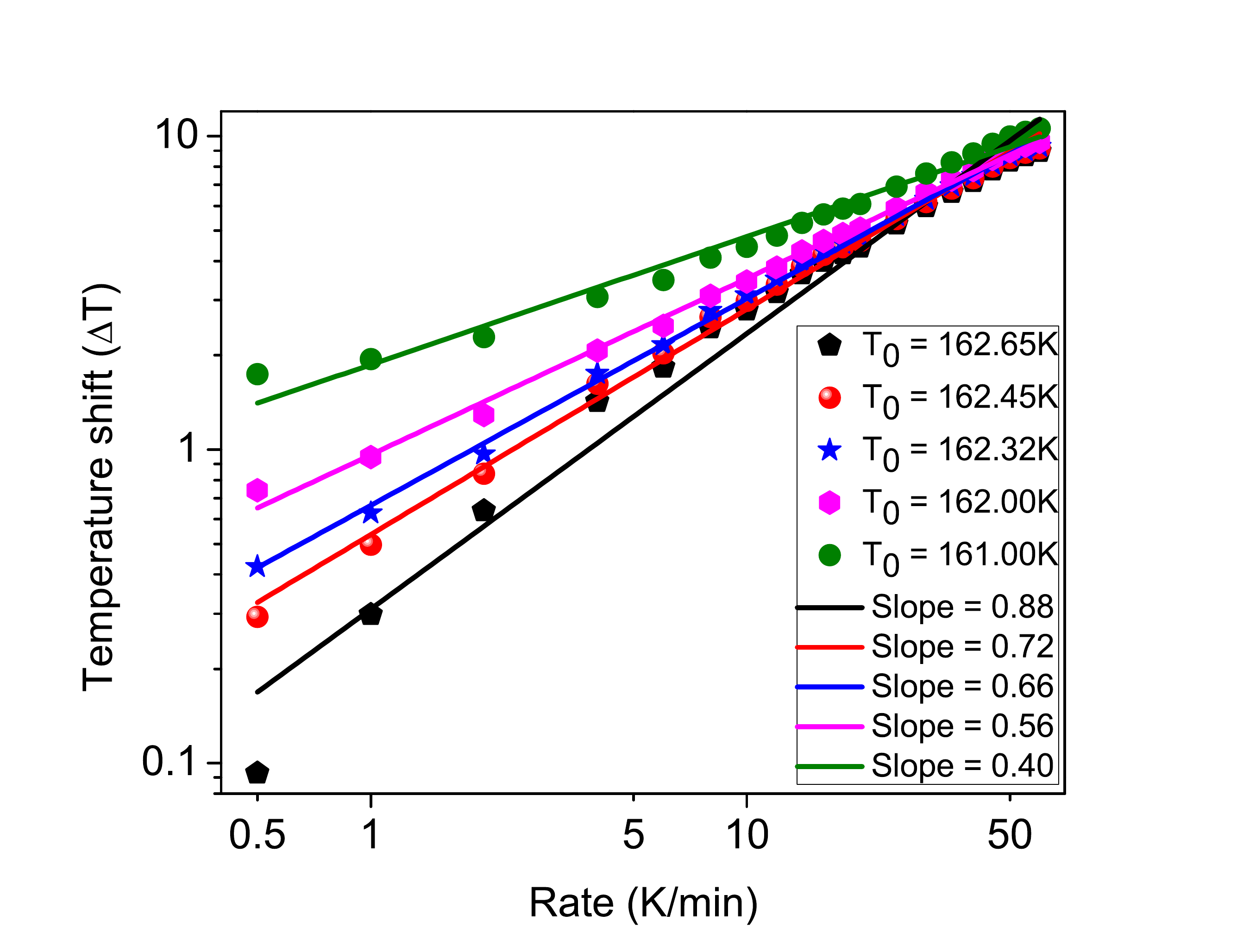}
\caption{Small differences in the value of $T_0$ can lead to different values of slope. Luckily one is able to obtain good fits only in a narrow window of slopes. }
\label{Figure_diffslopes}
\center
\end{figure}

\begin{figure}[h!]\label{Fig_ErrSlope}
\center
\includegraphics[scale=0.8]{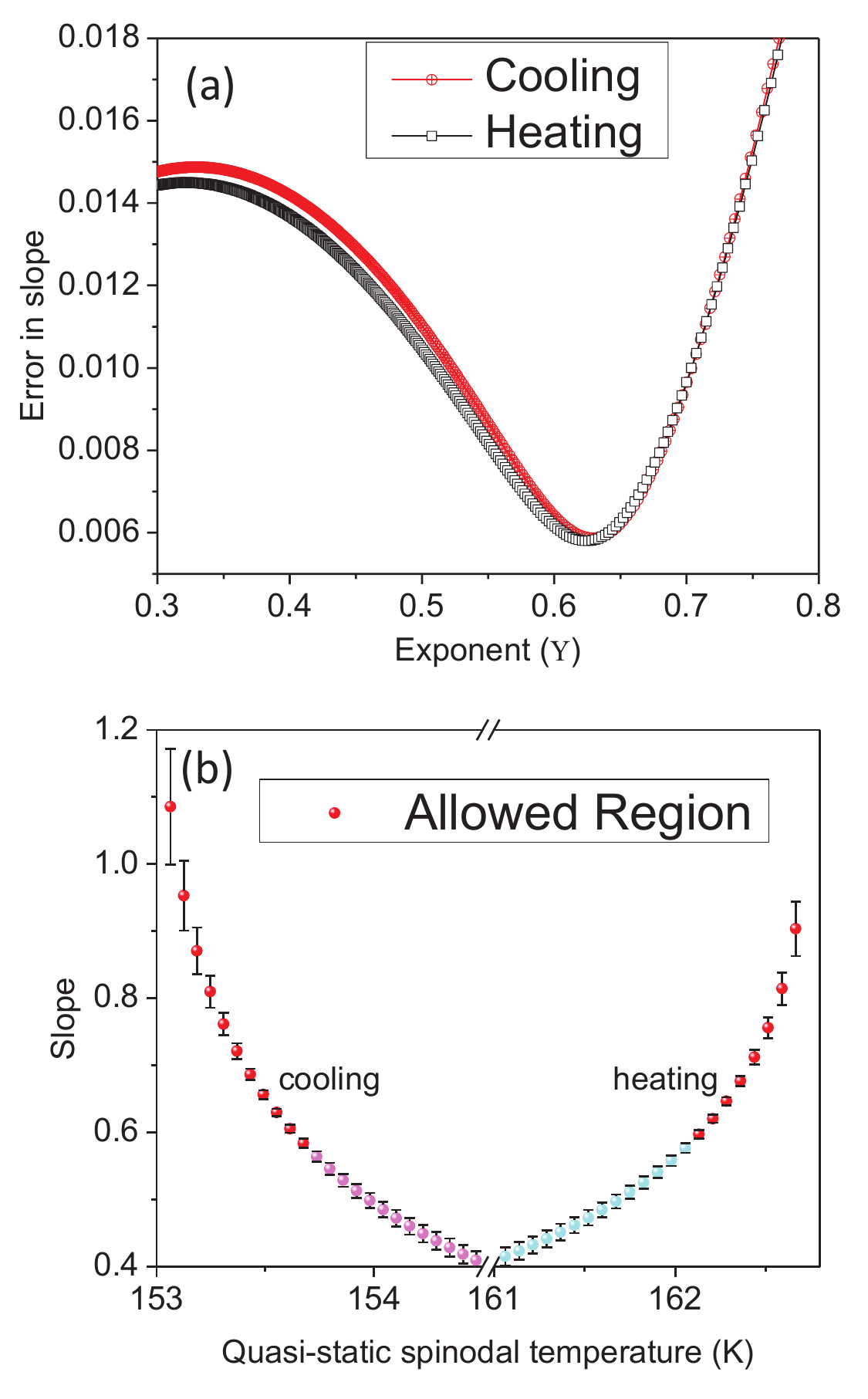}
\caption{(a) Error in the slope (on fitting Eq. \ref{eq1} to a straight line on log-log scale) {\it vs} the exponent $\Upsilon$ for different values of the transition temperature $T_0$. The minimum error is for $\Gamma\approx 0.62$, for both cooling and heating.  (b) The same as (a), except that the transition temperature $T_0$ is also plotted with the physically allowed values of the quasistatic spinodal temperature depicted in red. The error in slope  for each data point is also shown with error bars. The fact that $\Upsilon=0.62$ (where the error in slope is minimum) also corresponds to a permissible value of $T_0$ gives an independent consistency check.}
\label{Figure_errslopes}
\center
\end{figure}
\subsection{Method 2: Nonlinear fitting treating data points as independent quadruples}
For the experiment done at $N$ different ramp rates, let $T_i$ be the experimentally measured transition temperature at the ramp rate $R_i$,    $[i=1,..,N]$. We assume that the transition point $T_i$ (during heating) shifts with rate of change of temperature $R_i$, following a power law
\begin{equation} \label{eq1}
T_{i} = T_0 + aR_{i}^{\Upsilon}.
\end{equation}
$T_0$, the  transition temperature under quasi-static conditions and the prefactor $a$ are two unknown constants. If for another ramp rate, $R_j$, the measured transition temperature is $T_j$, then
\begin{equation} \label{eq2}
T_{j} = T_0 + aR_{j}^{\Upsilon}
\end{equation}
We eliminate $T_0$ by subtracting the above equations, i.e.,
\begin{equation}
(T_i - T_j) = a(R_{i}^{\Upsilon} - R_{j}^{\Upsilon}),\;\;i\neq j.
\end{equation}
There are $^NC_2=n_1$ (say) possibilities to pick out two values out of $N$. Furthermore, one can similarly eliminate $a$ by dividing an equation for a pair $\lbrace i, j\rbrace$ with the same equation for another pair $\lbrace k, l\rbrace$
\begin{equation}\label{eq:quadruples}
\frac{(T_i - T_j)}{(T_k - T_l)} = \frac{(R_{i}^{\Upsilon} - R_{j}^{\Upsilon})}{(R_{k}^{\Upsilon} - R_{l}^{\Upsilon})}
\end{equation}
Here $\lbrace i, j\rbrace\neq \lbrace k, l\rbrace$ but one should allow combinations such as $i= k$ (if $j\neq l$), etc. We thus have $^{n_1}C_2$ such transcendental equations which are numerically solved to get $^{n_1}C_2$ values of $\Upsilon$.

\begin{figure}[!h]\label{Fig_histSlope}
\begin{center}
\includegraphics[scale=0.4]{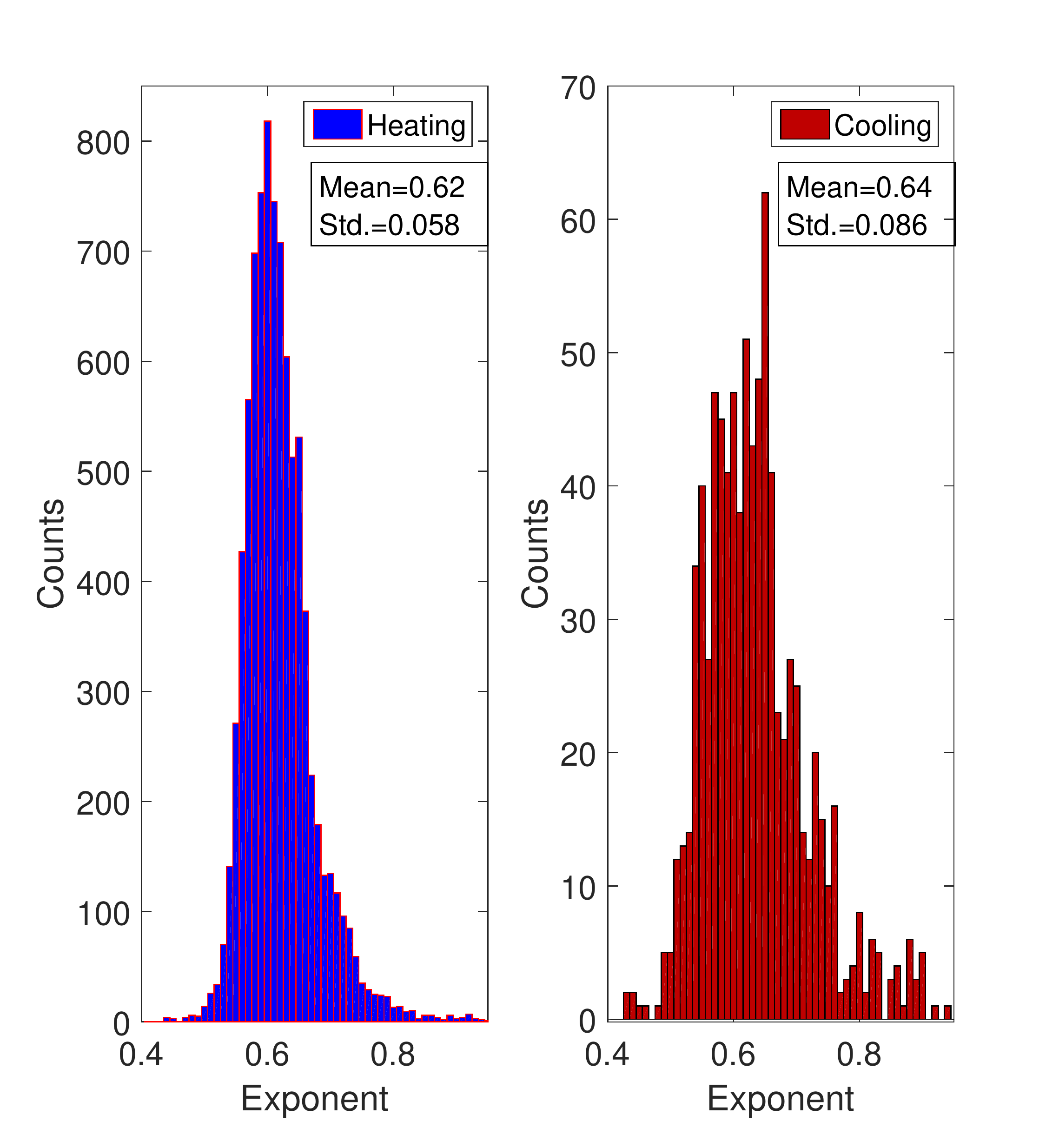}
\caption{An alternate method of determining the slope where each point is taken to be independent. Experimental constraints allowed for many more data points for the heating ramps and that is why the corresponding histogram has a smaller error.}
\end{center}
\end{figure}

Furthermore, the value of $\Upsilon$ so extracted gives a corresponding $T_0$, the quasistatic transition temperature, for the pairs $(\lbrace i, j\rbrace, \lbrace k, l\rbrace)$
\begin{eqnarray}
\nonumber T^{\lbrace i,j\rbrace|\lbrace k,l\rbrace}_0 = \frac{T_i - (\frac{R_i}{R_j})^\Upsilon T_j}{1-(\frac{R_i}{R_j})^\Upsilon}, \\  T^{\lbrace k,l\rbrace|\lbrace i,j\rbrace}_0 = \frac{T_k - (\frac{R_k}{R_l})^\Upsilon T_l}{1-(\frac{R_k}{R_l})^\Upsilon}.
 \end{eqnarray}

So far we have considered all pairs of points on equal footing and extracted out $^{n_1}C_2$ values of $\Upsilon$ and twice as many $T_0$'s. Physically, of course, $T_0$ must obey certain constraints; not every $T_0$ so obtained is acceptable.

In our experimental data and the theoretical model given by Eq. \ref{eq1} above, the transition temperature increases with the increase in heating ramp rate (and it decreases for cooling). Let $\delta T$ be the measured value of transition temperature difference ($|T_{R_1}-T_{R_2}|$) for two low rates (Let say $R_1 = 0.5$K/m and $R_2=1$K/m). Since Eq. \ref{eq1} implies that the difference in the transition temperatures should increase with the increase in heating rates, we can estimate two bounds on $T_0$ for it to be acceptable. Firstly, $T_0 < T_1$ where $T_1$ is the observed transition temperature for lowest heating rate. Secondly,  $(T_1-\delta T) < T_0$
The values of $\Upsilon$ which result in the inferred $T_0$ lying outside this window are rejected.
(Similarly we have the condition $(T_1-\delta T) > T_0 > T_1$  for the data taken under  cooling temperature ramps.)

Choosing $\delta T$ defined above to be $1$ K, we have plotted the histograms for the $\Upsilon$ for cooling and heating. The heating data has $20$ data points leading to $^{20}C_2=190$ pairs leading to $^{190}C_2=17955$ quadruples (used in Eq. \ref{eq:quadruples} above). $8592$ out of these yielded permissible values of $T_0$ and were used to plot the histogram [Fig. 15]. The cooling data had $11$ data points leading to $^{11}C_2=55$ pairs leading to $^{55}C_2=1485$ quadruples (used in Eq. \ref{eq:quadruples} above). Finally $861$ out of these were used to plot the histogram.

The mean values [$\Upsilon =0.62$ (heating) and $\Upsilon =0.64$ (cooling)] estimated by this method are in excellent agreement with the best fit estimate that we have calculated in the previous section. The standard deviation in the histograms are $0.058$ (heating) and $0.086$ (cooling) and can be used as estimates of error. Note that these errors are about an order of magnitude larger than the errors in slope mentioned in Fig. 14. Hence the values of the exponents are $\Upsilon=0.62\pm 0.06$ (heating) and $\Upsilon=0.64\pm 0.09$ (cooling)

\section{Compressible Ising model}
\subsection{Free energy in the mean field (Bragg-Williams) approximation}
Inclusion of `compressibility' into the Ising model was first suggested by Domb \cite{domb} and has been extensively studied  \cite{salinas}. The model is interesting because this simple extension of the Ising model produces a temperature-induced first-order transition and a tricritical point. Physically the model allows for a elementary description of structural phase transition.

Here we will derive the expression for the mean field free energy [Eq. 3 (main text)] that was used in the calculations shown in Fig. 4 (main text) and Fig, 5 (main text).

To fix the notation, let us start with the Hamiltonian for the usual Ising model
\begin{equation}\label{eq:0}
\mathcal{H}_I = -J_0\sum\limits_{(i,j)}\sigma_i\sigma_j.
\end{equation}
The sum is over nearest-neighbor sites of lattice i.e., $\sigma_i = \pm 1$, for all $i = 1,...,N$. In mean field Bragg-Williams approximation, the average internal energy $E =  <\mathcal{H}_I>$ is simply
\begin{equation}
E = -\frac{NqJ}{2}\phi^2.
\end{equation}
q is the coordination number of lattice and $\phi = \langle \sigma_i \rangle$ with $|\phi|\leq 1$ is the order parameter. The entropy for a system with $N$ lattice sites is
\begin{equation}
S(\phi) = Nk_B\{\ln2 - \frac{1}{2}[(1+\phi)\ln(1+\phi)+(1-\phi)\ln(1-\phi)]\}.
\end{equation}
Thus one obtains the well-known Bragg-Williams free energy \cite{chaikin-lubensky}
\begin{eqnarray}
\nonumber {\mathcal{F}_{Is}\over N} =  {k_BT\over 2}  [(1+\phi)\ln(1+\phi)+(1-\phi)\ln(1-\phi)]\} \\ \nonumber  -{qJ_0\over 2}\phi^2 - k_BT\ln2.\\
\end{eqnarray}

Let us now assume that the lattice is no longer rigid and can get distorted due to `spin-lattice' interactions. The effect of the lattice distortion is included via an additional harmonic elastic energy $\propto \frac{N}{2}(v-v_0)^2$ (where $v_0$ is the equilibrium volume and $v$ is the volume after distortion). More interestingly, one further assumes that the lattice distortion also affects the spin-spin exchange interaction in the Ising model, i.e., $J_0\rightarrow J(v)$ with
\begin{equation}\label{eq:2}
J(v) = J_0 - J_1(v-v_0).
\end{equation}
With these two modifications, we arrive at the Hamiltonian $\mathcal{H}_{CI}$ for the {\em compressible Ising model},
\begin{equation}\label{eq:0}
\mathcal{H}_{CI} = -J(v)\sum\limits_{(i,j)}\sigma_i\sigma_j + \frac{\mathcal{K}N}{2}(v-v_0)^2.
\end{equation}
 $\mathcal{K}$ is a positive parameter related to the inverse of compressibility.

The mean field free energy per site of the compressible Ising model thus has the extra compressibility term and a modified $J$
\begin{eqnarray}\label{eq:11}
{\mathcal{F}_{CI}\over N}=
\nonumber {k_BT\over 2}  [(1+\phi)\ln(1+\phi)+(1-\phi)\ln(1-\phi)]\} \\ \nonumber  -{1\over 2}qJ(v)\phi^2 - k_BT\ln2 + \frac{1}{2} \mathcal{K} (v-v_0)^2\\
\end{eqnarray}
As the derivative of Eq. \ref{eq:11} with respect to volume strain $(v-v_0)$ must vanish in equilibrium, $(v-v_0) = -\frac{J_1q\phi^2}{2\mathcal{K}}$. By inserting this expression back to the Eq. \ref{eq:11} we obtain
\begin{eqnarray}\label{eq:13}
\nonumber {\mathcal{F}_{CI}\over N} = \frac{k_BT}{2}[(1+\phi)\ln (1+\phi) + (1-\phi)\ln (1-\phi)] \\ \nonumber
-\frac{J_1^2q^2}{8\mathcal{K}}\phi^4 -\frac{J_0q}{2}\phi^2- k_BT\ln2.\\
\end{eqnarray}
We drop the last term $k_BT\ln2$ of Eq. \ref{eq:13}; it carriers no dynamics as it is independent of $\phi$. Furthermore, we can rescale the free energy, $\mathcal{F}_{CI}/N\rightarrow f(\phi, T)$ so that there are only two free parameters, $T_c$ and $\xi$. We thus have the free energy in the form of Eq. 3 of the main text.
\begin{eqnarray}\label{eq:14}
\nonumber f(\phi, T) =  \frac{T}{2T_c}[(1+\phi)\ln (1+\phi) + (1-\phi)\ln (1-\phi)]\\ \nonumber - \xi \phi^4 -\frac{\phi^2}{2}\\
\end{eqnarray}
where $\xi = {J_1^2q\over 8J_0\mathcal{K}}$ and $T_c = {qJ_0\over k_B}$.

On Taylor expanding the entropy term, it is obvious that the $\phi^4$ term would become negative for ${T\over 12T_c}-\xi<0$. One would then observe a thermally induced APT. One spinodal is at the critical temperature $T_c$ and second spinodal depends on the value of $\xi$ [Fig. 16]. Such a Taylor expansion with a negative $\phi^4$ term also naturally leads to the free energy expression in the Landau form with a `$\phi^6$' term.

Finally, it is of interest to note that the role of lattice compressibility in the Mott transition {\em around the critical point} has been discussed by making an extension to the Hubbard model \cite{hassan} which is similar in spirit to the above discussion. Instability at half filling in Hubbard model with respect to lattice contraction \cite{hassan} yields a strongly first order phase transition.

\begin{figure}[h!]\label{Fig_FOPT_line}
\center
\includegraphics[scale=0.25]{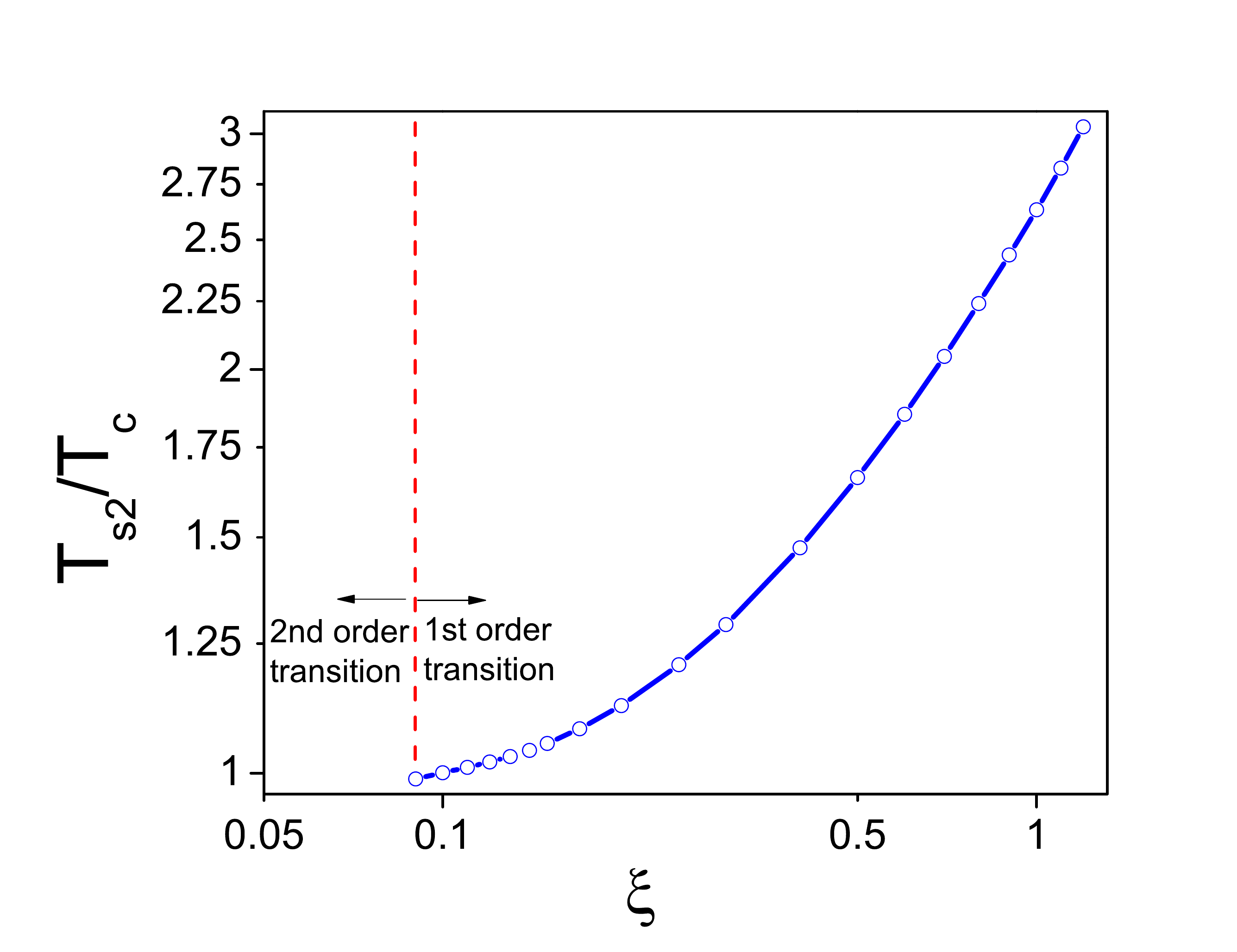}
\caption{Calculated phase diagram of the compressible Ising model where the value of the second spinodal temperature $T_{s2}$ (when it exists) is plotted in units of the critical temperature $T_c$ {\em vs} the parameter $\xi$.  The other spinodal temperature $T_{s1}=T_c$. There is no first-order transition for $\xi<1/12$. If the metastable phase persists up to the spinodals, the width of thermal hysteresis $\Delta T=(T_{s2}-T_c)$ approaches zero at tricriticality and monotonically increases with $J_1$.}
\label{FORT}
\center
\end{figure}

\subsection{Dynamic hysteresis}
In Fig. 17, we have plotted the time evolution of the order parameter $\phi$ under a linear ramp in temperature at rates varying from $0.2$ K/min to $50$ K/min. Fig. 17 (a) is generated using Eq. 1 (main text) with the form of the free energy given by Eq. 19. Fig. 17 (b) indeed shows that the  transition temperature does indeed dynamically shift with the exponent $\Upsilon=2/3$.

\begin{figure}[h!]\label{Fig_TDGL_DynamicHyst}
\center
\includegraphics[scale=0.35]{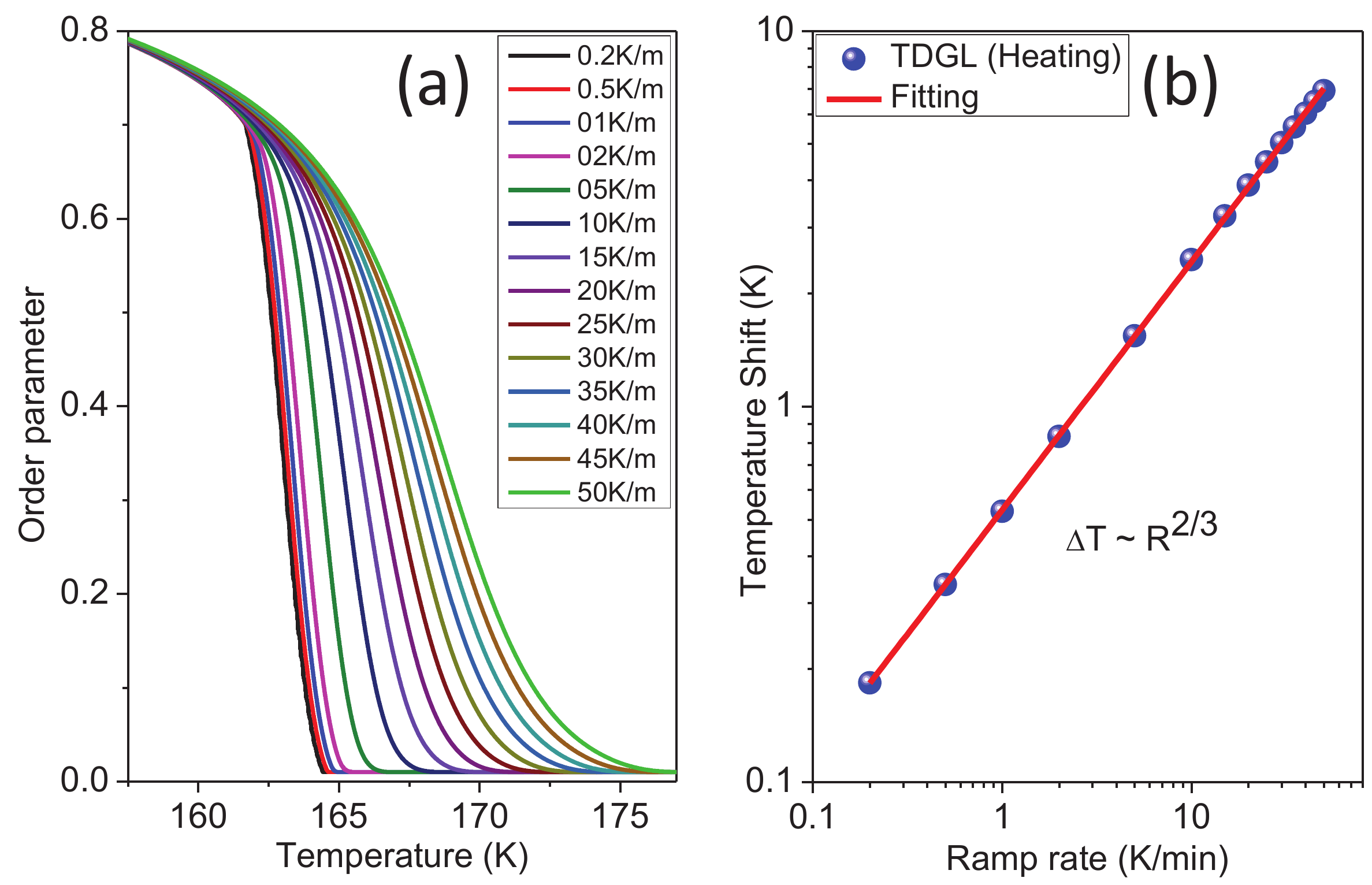}
\caption{(a) Temporal evolution of the order parameter with the deterministic TDL equation with the free energy given by the compressible Ising model with the temperature linearly increasing at different rates.  (b) Shift in the transition temperature {\em vs} temperature scan rate.}
\label{FORT}
\center
\end{figure}

\section{Dynamic hysteresis and its analogy with finite size scaling}
We have seen in Fig. 17 above that $\Upsilon=2/3$ emerges from the numerical solution of the time dependent Landau equation (Eq. 1 [main text]). This is direct evidence of barrier-free evolution (continuous ordering) around spinodal-like instabilies. In this final section, we give a heuristic explanation for dynamic scaling via analogy with finite size scaling \cite{zhong_arxiv2015, zhong_prl2005}.

In the theory of equilibrium critical phenomena for continuous transitions, the divergence of the correlation length captures the singular behavior of all the thermodynamic variables close to the critical temperature $T_c$. At the critical point, as the correlation length diverges, the system becomes scale-free and various physical quantities show power-law scaling \cite{chaikin-lubensky}. This scaling picture extends also to critical dynamics via the ansatz that the characteristic time scale $\tau$ also diverges (critical slowing down).

\begin{figure}\label{fig:FSS_tempshift}
\begin{center}
\includegraphics[scale=0.5]{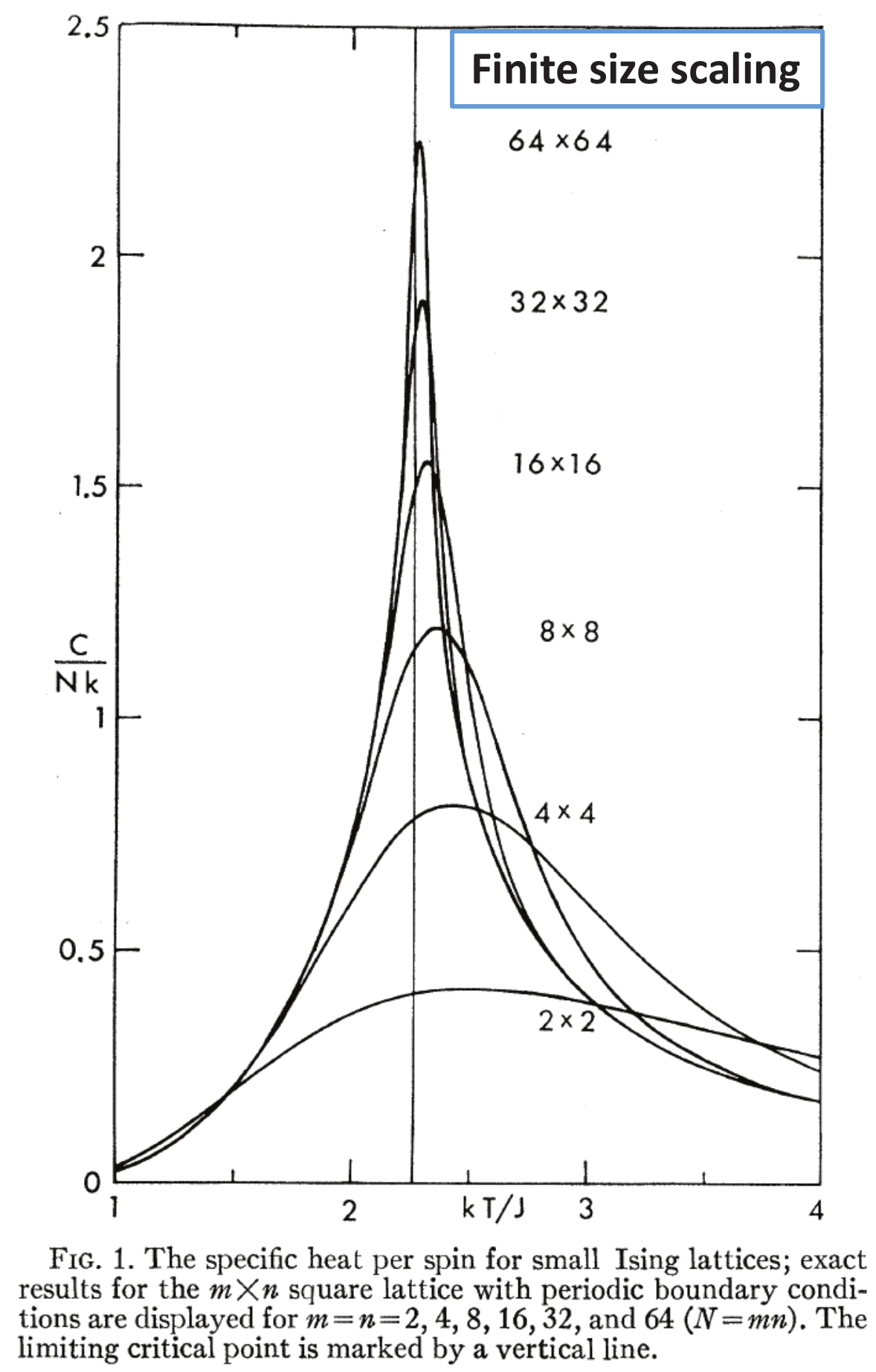}
\caption{Figure from classic reference on finite-size-scaling by Ferdinand and Fisher \cite{Ferdinand}.  A shift in the transition temperature, accompanied by broadening in the transition is seen as the sample size is reduced from $64\times 64$ to $2\times 2$ sites.}
\end{center}
\end{figure}
When one is dealing with a finite system (as in a simulation, but the argument is also valid for a real finite-sized system \cite{Henkel}), the correlation length  cannot diverge but is bounded by the system size $L$. One can modify the scaling laws to account for a finite system, using the finite-size scaling prescription \cite{Ferdinand, Henkel}. An important attribute of finite-size scaling is that there is a systematic shift in the transition temperature with the system size. For example, the peak of the specific heat scales with the system size $L$ as \cite{Henkel}
\begin{equation}\label{eq:shift exponent}
|\Delta T|\sim |T_c-T_c(L)|\sim L^{-\lambda} ,
\end{equation}
where $T_c$ is the transition temperature for an infinite system, and $T_c(L)$ is the transition temperature for a system of size $L$. $\lambda$ is referred to as the shift exponent. Fig. 18 shows how the transition temperature and the width of the specific heat peak vary with the system size \cite{Ferdinand}.

One can take an analogous view for the dynamical shift in the transition temperature. Since we are dealing with metastable states, any description must go beyond equilibrium and directly address the kinetics of phase ordering. A spinodal-like singularity would imply a (critical-like) slowing down due to the divergence of the characteristic response time of the system. As the temperature is being linearly swept across the spinodal, the sluggish response of the system at the spinodal causes an overshoot in the transition. The fact that there is a power-law scaling would further suggest some scale-free behavior due to this divergence.

In the work of Zhong \cite{zhong_arxiv2015, zhong_prl2005}, these ideas have been made more precise by mapping this problem to the finite-size scaling scenario discussed above. The fact that the system under the action of continuously varying temperature has a finite time to respond can be formulated as a ``finite time scaling" problem \cite{zhong_arxiv2015, zhong_prl2005}. If one assumes that such a spinodal instability exists (the whole argument hinges on this assumption), it is possible to get a scaling expression analogous to Eq.~\ref{eq:shift exponent} above with the system size $L$ replaced by $R$, the rate of change of the control parameter (field or temperature) under linear sweep, $|\Delta T|\sim |T_c-T_c(R)|\sim R^{\Upsilon}$. The exponent $\Upsilon=2/3$ is obtained within mean field theory \cite{zhong_arxiv2015, zhong_prl2005}.

\begin{center}
{\bf REFERENCES}
\end{center}

References appearing in the Supplemental Material are in the common references section at the end of the main text (page 4-6).

\begin{references}
\bibitem{footnote_otherExponents}The scaling ansatz for dynamic hysteresis under non-deterministic evolution \cite{Chakrabarti_rmp}, does not seem to have similar microscopic justification.  A range of exponents (very different from $\Upsilon=2/3$) have been empirically observed \cite{rao,Sengupta-Marathe-Puri, nematic-smectic, He-Wang} in simulations and experiments..
\bibitem{footnote zeroth-order}Since the free energy itself is discontinuous. 
\end{references}
\end{document}